\documentclass[epj,final]{svjour}
\usepackage{graphicx}
\newcommand{\beq}{\begin{equation}}
\newcommand{\beqa}{\begin{eqnarray}}
\newcommand{\eeq}{\end{equation}}
\newcommand{\eeqa}{\end{eqnarray}}
\newcommand{\abs}[1]{\left\vert#1\right\vert}
\newcommand{\abslnexdy}{\vert\!\ln x_{\rm dyn}\vert}
\newcommand{\bin}[2]{{#1\choose#2}}
\renewcommand{\d}{{\rm d}}
\newcommand{\dpar}{\partial}
\newcommand{\ds}{\displaystyle}
\newcommand{\e}{{\rm e}}
\newcommand{\eps}{\varepsilon}
\newcommand{\exdy}{x_{\rm dyn}}
\newcommand{\exi}{x_{\rm int}}

\renewcommand{\frac}[2]{\displaystyle{\displaystyle#1\over\displaystyle#2}}
\newcommand{\half}{{\textstyle{\frac12}}}
\newcommand{\lra}{\Longrightarrow}
\renewcommand{\max}{{\rm max}}
\newcommand{\mean}[1]{\langle#1\rangle}
\renewcommand{\o}{\omega}
\newcommand{\prob}{\mathop{\rm Prob}\nolimits}
\newcommand{\quarter}{{\textstyle{\frac14}}}
\newcommand{\react}[2]{\mathrel{\mathop{#1}\limits_{}^{#2}}}
\newcommand{\s}{\sigma}
\newcommand{\st}{\;|\;}
\newcommand{\sign}{\mathop{\rm sign}\nolimits}
\newcommand{\var}{\mathop{\rm var}\nolimits}
\newcommand{\w}{\widehat}
\newcommand{\xidy}{\xi_{\rm dyn}}
\newcommand{\xii}{\xi_{\rm int}}
\newcommand{\xiic}{\xi_{{\rm int,}c}}
\newcommand{\xiione}{\xi_{{\rm int,}1}}
\newcommand{\C}{{\cal C}}
\newcommand{\E}{{\cal E}}
\renewcommand{\L}{{\cal L}}
\newcommand{\M}{{\bf M}}
\renewcommand{\S}{\Sigma}
\newcommand{\T}{\Gamma}
\begin{document}
\title{Dynamical diversity and metastability in a hindered granular column
near jamming}
\author{J.M. Luck\inst{1} \and Anita Mehta\inst{2}}
\institute{
Service de Physique Th\'eorique\thanks{URA 2306 of CNRS},
CEA Saclay, 91191 Gif-sur-Yvette cedex, France.
\email{jean-marc.luck@cea.fr}
\and
The Radcliffe Institute for Advanced Study, Harvard University,
34 Concord Avenue, Cambridge, MA 02138, USA.
\email{anita.mehta@cea.fr}}
\date{}
\abstract{
Granular media jam into a panoply of metastable states.
The way in which these states are achieved depends on the nature
of local and global constraints on grains;
here we investigate this issue by means of a non-equilibrium stochastic model
of a hindered granular column near its jamming limit.
Grains feel the constraints of grains
above and below them differently, depending on their position.
A rich phase diagram with four dynamical phases
(ballistic, activated, logarithmic and glassy) is revealed.
The statistics of the jamming time and of the metastable states
reached as attractors of the zero-temperature dynamics
is investigated in each of these phases.
Of particular interest is the glassy phase, where intermittency
and a strong deviation from Edwards' flatness are manifest.}
\maketitle

\section{Introduction}
\setcounter{equation}{0}
\def\theequation{1.\arabic{equation}}

Granular media \cite{0} are by now recognised as
being paradigms of complexity \cite{comp}, especially near their
jamming limit \cite{sidjamming}.
The fact that most grains are too large
to be perturbed by the effect of room temperature leads to an `athermal'
dynamics, which is a major cause of this complexity -- configurations
once generated, are remembered, and their hysteretic effects persist in
any ensuing dynamics, and the new configurations generated therefrom.
Another
origin of complexity in such systems is their generic disorder; this leads
in particular to a random landscape of metastable states being explored if
a granular system is suitably `quenched', rather than a crystallisation into
an appropriate ordered state.
The way in which this landscape is explored depends
strongly on the driving forces applied, in the absence of any real
thermodynamics.

The work we present here is an attempt to explore many of these issues, in
particular those to do with the nature of ground states in a system near
jamming, and how these are reached.
This is the main motivation for our
focusing on the effect of `zero-temperature' dynamics on the model we will
later introduce, since zero-temperature dynamics are known to be the route
to systemic ground states.
Another aspect of interest around which our model
was designed was the exploration of spatial inhomogeneities, to answer
questions such as: how does position along a column of grains influence the
dynamics observed there? This is an important question for several reasons,
one of the earliest being that experimental measurements of density
along a column of grains
revealed wide variations \cite{sid}, depending on whether they were near
the top, the bottom or the middle -- in its turn, this implied that the
compactivity of the system \cite{edwards} was non-uniform.
A more visible manifestation of spatial inhomogeneities is the presence of
force chains \cite{bob}
and bridge networks \cite{bridges}, which are unique signatures of the
granular state.

Given the complexity of issues we wish to investigate, we have chosen
a minimal model to work with, whose ingredients are based on our experience
with several earlier ones \cite{I,II,III,IV}.
A feature that all the models
share is that of orientational disorder; every grain is allowed to occupy
one of two states, corresponding to `ordered' and `disordered'.
Disordered
orientations generate voids and waste space, whereas ordered ones do not.
Implicit in
this description is the effect of shape, which is most easily understood
in terms of the rectangular grains of aspect ratio $a$
considered in \cite{I,II}.
Grains aligned
along their long edges (length $1$) result in a fully packed column,
whereas those perched
on their short edges (length $a$) leave voids of size $1-a$.
The horizontal orientation is thus ordered, and the vertical one disordered.

Our models became progressively more sophisticated.
The earliest model of rectangular grains considered in \cite{I,II}
was strictly non-interacting, with the only effects included being due
to gravity and
excluded volume -- grains could not overlap, and those that were deeper
in a column were less free to move.
One of the major extensions in \cite{III,IV} was the introduction of
the shape parameter $\eps$ to include grains of
arbitrary, i.e., non-rectangular shape -- such grains can have multiply many
orientations in reality, but the two-state description was retained
in the interests of simplicity.
The parameter $\eps$ was allowed
to vary over all rational and irrational numbers, with a view to describing
regular and irregular grains, as will be explained further below.

Another difference between the two models was that the first one \cite{I,II}
interpolated between
jammed and fluidised regimes, whereas the second one was explicitly
constructed to examine the jamming regime.
In the latter case, it seemed reasonable to
assume that translational diffusion was
essentially absent, with all compaction occurring via orientational
rearrangements -- it was therefore appropriate to focus on a column, rather
a box \cite{I,II}, of grains.
Interactions were introduced in the column model of \cite{III,IV}
such that every grain now not only
felt the weight of grains above it, but was also constrained by
all of their orientations.
This modelling was appropriate as a representation
of grains in the free surface layer, which were too far from the base to
feel an undertow; in the jamming limit, however, such grains would be
expected to feel the orientational constraints arising from grains
{\it above} them.

The present model incorporates all of the above features, and generalises
them to include the presence of
orientational constraints arising from grains {\it below} a given grain.
Unlike our previous models \cite{I,II,III,IV}, where interactions propagated
downwards from the top, here they propagate {\it both upwards and downwards}.
Clearly the extent of this propagation depends on grain position
In this sense we model a column of grains with a top, a middle and a bottom.

We devote this paper to a comprehensive investigation of the ground states
of this model, and how they are reached via zero-temperature dynamics.
The main questions we will answer along the way will be related to several of
the issues mentioned above; in particular the issue of spatial inhomogeneities
will be relevant, since the dynamical regimes attainable via this model
will depend on which part of the column -- top, middle or bottom -- is being
examined.
Additionally, we will see that there is a panoply of {\it metastable}
ground states available to the system; the dynamics of their attainment
will allow us to classify the appropriate regimes as {\it ballistic},
{\it logarithmic}, {\it activated} and {\it glassy}.
The glassy regime is by far the most novel and
interesting of these regimes, and will be investigated in
greater depth than the others; its full exploration is, however,
reserved for future work.

The plan of this paper is as follows.
The definition of the model is given in Section 2.
Section 3 contains an investigation of its static properties,
with an emphasis on the ground states.
Section 4 presents a study of zero-temperature dynamics:
a rich phase diagram with four dynamical phases is revealed
and investigated thoroughly.
A discussion is presented in Section 5.
Exact results for small systems, as well as other technical results,
are presented in three appendices.

\section{The model}
\setcounter{equation}{0}
\def\theequation{2.\arabic{equation}}

Like the unhindered (fully directional) model described in \cite{III,IV},
the present model consists of a finite column of $N$ grains,
labelled by their depth $n=1,\dots,N$.
Each grain assumes two orientational states,
which are referred to as {\it ordered} and {\it disordered}.
We set $\s_n=+1$ (resp. $\s_n=-1$) if grain number $n$
is ordered (resp. disordered).
A configuration of the column is therefore uniquely defined
by the orientation variables $\{\s_n\}$.
There are $2^N$ such configurations.

The model is defined by a stochastic dynamics
which do not obey detailed balance.
In the present context, detailed balance essentially means a symmetry
in $m$ and $n$ on the dynamical effect of
grain orientation $\s_m$ on $\s_n$.
The expressions for the local fields (\ref{hdef}), (\ref{hjdef})
clearly do not obey such `action and reaction'.
Our model is therefore intrinsically out of equilibrium,
and its stationary state at finite temperature
is a genuine {\it non-equilibrium steady state}

More precisely, the model is defined as follows.
Grains are selected in a random sequential fashion
and updated with the orientation-flipping rates
\beq
\left\{\matrix{
w(\s_n=+1\to\s_n=-1)=\exp\left(-\frac{\lambda_n+H_n}{\T}\right),\hfill\cr\cr
w(\s_n=-1\to\s_n=+1)=\exp\left(-\frac{\lambda_n-H_n}{\T}\right),\hfill
}\right.
\label{rates}
\eeq
where, along the lines of previous work \cite{I,II,III,IV}:

\noindent$\bullet$
$\T$ is a dimensionless vibration intensity, referred to as temperature,
and related to the `fast' temperature \cite{0} in granular media.

\noindent$\bullet$
$\lambda_n$ is the activation energy felt by grain $n$.
We make the assumption that it increases linearly with the depth $n$,
but otherwise does not depend on grain orientations.
We set
\beq
\lambda_n=\frac{n\T}{\xidy},
\eeq
so that the local frequency
\beq
\o_n=\exp\left(-\frac{\lambda_n}{\T}\right)=\exp\left(-\frac{n}{\xidy}\right)
\label{odef}
\eeq
falls off exponentially, with a characteristic length $\xidy$.
This {\it dynamical length} corresponds to the depth of the boundary layer
beyond which grains are frozen out by the sheer weight of grains above them.

\noindent$\bullet$
$H_n$ is the local ordering field felt by grain $n$,
which determines the orientational response of grain $n$
to the orientations $\{\s_m\}$ of all the other grains.
In previous work \cite{III,IV} the local field $H_n$ only involved
the uniform effect of the upper grains ($m=1,\dots,n-1$).
In the present model, we also take into account the {\it back-propagation}
from grains below a given grain $n$ ($m=n+1,\dots,N$).
This effect cannot be similarly uniform.
We assume for simplicity that upward interactions are exponentially damped,
with a characteristic length $\xii$, the {\it interaction length}.
We therefore set
\beq
H_n=h_n+gj_n,
\label{hdef}
\eeq
where we denote by $h_n$ the {\it uniform} effect of grains {\it above} $n$
($m=1,\dots,n-1$),
and by $j_n$, the {\it non-uniform} effect of grains {\it below} $n$
($m=n+1,\dots,N$), whose strength is measured by a small (positive)
coupling constant $g$.
Both components $h_n$ and $j_n$ of the local field
contain the contribution of every single grain orientation $\s_m=\pm1$
through the quantity $f(\s_m)$.
As before \cite{III,IV}, the latter assumes the following two values:
\beq
f(\s_n)=\left\{\matrix{
\eps\hfill&\hbox{if}\hfill&\s_n=-1,\cr
-1\ \hfill&\hbox{if}\hfill&\s_n=+1.}\right.
\label{fdef}
\eeq
A useful equivalent formula is the following:
\beq
f(\s_n)=\half(\eps-1-(\eps+1)\s_n).
\label{fdefalt}
\eeq
In terms of this quantity, $h_n$ and $j_n$ are:
\beq
\matrix{
\ds{h_n=\sum_{m=1}^{n-1}f(\s_m)},\hfill\cr
\ds{j_n=\sum_{m=n+1}^Nf(\s_m)\,\exp\left(-\frac{m-n}{\xii}\right)}.}
\label{hjdef}
\eeq

The positive shape parameter $\eps$
represents an `effective aspect ratio' for a grain of arbitrary shape.
This interpretation originated in the framework of the non-interacting model
\cite{I,II} with rectangular grains.
Rational values of $\eps=p/q$ imply that the grain size is expressible
by a rectangle of sides $p$ and $q$.
Such grains are brick-like, and therefore can be packed perfectly
to build some periodic tiling.
On the other hand, when $\eps$ is irrational, such tilings cannot be built,
so that the most close-packed
states are those of optimal, rather than perfect packing.
We thus continue to make the following equivalence \cite{III,IV}:
{\it rational} values of $\eps$ imply grains of {\it regular} shape,
while {\it irrational} values of $\eps$ imply grains of {\it irregular} shape.

The parameters of the model are therefore the number of grains $N$,
the coupling constant $g$, the shape parameter $\eps$,
and most importantly the interaction and dynamical lengths $\xii$ and $\xidy$.
The unhindered model of references \cite{II,III,IV}
is recovered in the absence of coupling ($g=0$).
Throughout the following, we use the notations
\beq
\exdy=\e^{-1/\xidy},\quad\exi=\e^{-1/\xii}.
\label{zdef}
\eeq

To close up, we mention the following recursion relations obeyed
by the components $h_n$ and $j_n$:
\beq
\matrix{
h_n=h_{n-1}+f(\s_{n-1}),\hfill\cr
j_n=\exi\left(f(\s_{n+1})+j_{n+1}\right).}
\label{hjrel}
\eeq
These relations, together with the boundary values $h_1=j_N=0$,
provide a fast algorithm to evaluate the local fields,
to be used in numerical simulations.
On the other hand, the relations (\ref{hjrel}) also imply
\beq
f(\s_n)=h_{n+1}-h_n=\frac{j_{n-1}}{\exi}-j_n.
\eeq
The latter equation determines the $j_n$ in terms of the $h_n$:
\beq
\matrix{
j_n=-\exi h_{n+1}+j_n^{(1)},\hfill\cr
\ds{j_n^{(1)}=(1-\exi)\sum_{m=n+2}^N\exi^{m-n-1}h_m+\exi^{N-n}h_{N+1}}.}
\label{jexplicit}
\eeq

Throughout the following, we choose to impose for convenience
the boundary condition that the uppermost grain is ordered:
\beq
\s_1=+1.
\label{init}
\eeq

\section{Statics. Ground states}
\setcounter{equation}{0}
\def\theequation{3.\arabic{equation}}

The dynamical rules simplify in the zero-temperature limit ($\T\to0$).
Indeed (\ref{rates}) yields (provided $H_n\ne0$):
\beq
\matrix{
\frac{w(\s_n=-1\to\s_n=+1)}{w(\s_n=+1\to\s_n=-1)}
=\exp\left(\frac{2H_n}{\T}\right)\hfill\cr
{\hskip 37.6truemm}
\to\left\{\matrix{\infty\hfill&\hbox{if}\hfill&H_n>0,\hfill\cr
0\hfill&\hbox{if}\hfill&H_n<0.\hfill
}\right.
}
\label{zero}
\eeq

Along the lines of references \cite{II,III,IV},
a {\it ground state} of the column is defined as a configuration
where the orientation of every grain is aligned along its local field:
\beq
\s_n=\sign H_n=\left\{\matrix{
+1\ \hfill&\hbox{if}\hfill&H_n>0,\hfill\cr
-1\hfill&\hbox{if}\hfill&H_n<0.\hfill
}\right.
\label{zerost}
\eeq

The ground states of the unhindered model ($g=0$)
have been investigated in \cite{III,IV}.
In that case, the local field $H_n=h_n$ acting on grain $n$
only depends on the grains above $n$.
Equation (\ref{zerost}) therefore yields
a recursive procedure allowing one to construct ground states:
\beq
\left\{\matrix{
h_n>0\lra\s_n=+1,\ \hfill&h_{n+1}=h_n-1,\hfill\cr
h_n<0\lra\s_n=-1,\ \hfill&h_{n+1}=h_n+\eps.\hfill
}\right.
\label{step}
\eeq
In a ground state, all the local fields $h_n$ lie in the range
\beq
-1\leq h_n\leq\eps.
\label{range}
\eeq

For the present model with a non-zero coupling constant $g$,
things are more complex.
The local field $H_n$ given in (\ref{hdef}) now depends both
on the grains above $n$ (through $h_n$)
and on the grains below $n$ (through $j_n$).
The condition (\ref{zerost}) therefore couples
all the orientation degrees of freedom.
In particular the ground states admit no recursive construction
similar to (\ref{step}).
It is therefore a non-trivial task to generate all the ground states
of a finite column of $N$ grains, whose number depends on
$g$ and $\xii$ in general.

The situation however simplifies in the weak-coupling regime ($g\ll1$),
where the ground states can be understood
in terms of those of the unhindered model ($g=0$)
by means of a stability analysis.
Just as in the case of the unhindered model \cite{III,IV},
rational and irrational values of $\eps$ have to be considered separately,
in the case of the present hindered model.

\subsection{Rational $\eps$ (regular grains)}

We first recall some facts about the ground state structure
in the unhindered model.
Using (\ref{step}) in that case, we find that
whenever the shape parameter $\eps$ is a rational number (in irreducible form)
\beq
\eps=\frac{p}{q},
\eeq
the fields $h_n$
vanish at all depths $n$ such that $n-1$ is an integer multiple of $p+q$.
The corresponding orientation $\s_n$ is left unspecified, and can be chosen at
will.
Ground states are therefore all the random sequences
made of two well-defined patterns, $P_1=+-\cdots$ and $P_2=-+\cdots$
These patterns have the same length $p+q$;
they are made of $p$ ordered and $q$ disordered grains,
and only differ by the orientations of their two uppermost grains.
The model therefore has an exponentially large
number of ground states, of the form $\exp(N\S)$,
where the configurational entropy \cite{sconf} reads
\beq
\S=\frac{\ln 2}{p+q}.
\eeq

It turns out that these ground states
are still the exact ones for the present model,
as long as the coupling constant $g$ is smaller than some threshold $g_c$.
In order to show this, we approach the problem via a stability analysis.
We assume that the $h_n$ field predominates,
consider $j_n$ as a small perturbation,
and see whether the ground states for $g=0$ are still stable
in the presence of the additional feedback of the $j_n$ field.

\noindent$\bullet$
Consider first the case when a grain is at
a depth $n=(p+q)m+1$ for some integer $m=1,2,\dots$
The local field $h_n$ vanishes, so that the orientation $\s_n=\eta=\pm1$
can be chosen at will for $g=0$, i.e., in the zeroth order approximation.
Once this choice is made, we have $h_{n+1}=f(\eta)$,
$\s_{n+1}=-\eta$, and $h_{n+2}=\eps-1$.
In order to see if the ground state is stable when the effect
of the $j_n$ field is included, we have to first estimate $j_n$.
Equation (\ref{jexplicit}) can be iterated once, yielding
\beq
\matrix{
j_n=-\exi h_{n+1}+\exi(1-\exi)h_{n+2}+j_n^{(2)},\hfill\cr
\ds{j_n^{(2)}=(1-\exi)\sum_{m=n+3}^N\exi^{m-n-1}h_m+\exi^{N-n}h_{N+1}}.}
\eeq
First, we notice that the inequalities (\ref{range})
allow one to bound the remainder $j_n^{(2)}$
as $-\exi^2\leq j_n^{(2)}\leq\exi^2\eps$.
Then, the values of $h_{n+1}$ and $h_{n+2}$ obtained above
yield the inequalities
$j_n>\exi(1-\exi)\eps$ if $\s_n=+1$ and $j_n<-\exi(1-\exi)$ if $\s_n=-1$.
Therefore, the orientation $\s_n$ is aligned with the local field $H_n=gj_n$,
irrespective of the coupling constant $g$.

\noindent$\bullet$
Consider now a depth $n$ which is not of the form $n=(p+q)m+1$, so that
the local field $h_n$ is non-vanishing.
As $h_n$ is an integer linear combination
of terms $f(\s_m)$ equal either to $-1$ or to $\eps=p/q$,
it therefore obeys
\beq
\abs{h_n}\ge\frac{1}{q}.
\label{hest}
\eeq
On the other hand, (\ref{jexplicit}) can again be used to estimate $j_n$.
The inequalities (\ref{range}) imply $-\exi\leq j_n^{(1)}\leq\exi\eps$, and
\beq
\abs{j_n}<(1+\eps)\exi=\frac{p+q}{q}\,\exi.
\label{jest}
\eeq
The inequalities (\ref{hest}) and (\ref{jest}) imply
\beq
\abs{\frac{h_n}{j_n}}>\frac{1}{(p+q)\exi}.
\label{hjest}
\eeq

The full local field $H_n$ therefore
has the same sign as $h_n$ for all $n$, provided the coupling constant $g$
is smaller than some threshold $g_c$,
so that $h_n$ is the dominant field in this weak-coupling regime.
While the arguments above do not allow one to predict exact values
of $g_c$\footnote{The exact threshold coupling $g_c$ will, however,
be determined later on for $\eps=1$ (see (\ref{gc1})).}
for generic rational $\eps$, they do yield a lower bound:
\beq
g_c\ge\frac{1}{(p+q)\exi}.
\label{gcdef}
\eeq

Also, similar arguments show that no other ground states exist in this model
in the same range of $g$, allowing us to identify, for all $g<g_c$,
the ground states of this model with those of the
unhindered model \cite{III,IV}.

\subsection{Irrational $\eps$ (irregular grains)}

Once again, we review the nature of the ground states
for irregular grains (irrational $\eps$) in the earlier
unhindered model \cite{III,IV}.
A unique ground state
(corresponding to optimal, rather than perfect packing) is obtained
for each value of $\eps$ in that case, such that
all the fields $h_n$ generated by the recursion procedure (\ref{step})
are non-zero.
The main feature of this ground states is that it is {\it quasiperiodic}.

The nature of the ground states in the hindered model under discussion here
can be predicted by analogy with the low-temperature excitations
of the unhindered model.
Indeed it turns out that the presence of
a weak coupling ($g\ll1$) nucleates disorder
in this model, in the same way as a low but finite temperature
($\T\ll1$) destroys the ground state of the unhindered model \cite{IV}.

More precisely, as long as the local field $H_n$ has the same sign as $h_n$,
i.e., for $\abs{h_n}>g\abs{j_n}$, the ground states of both unhindered
and hindered models are identical.
The depth up to which this stability condition is satisfied
can be estimated as follows.
Equation (\ref{jest}) shows that typical values of $j_n$ are of order $\exi$.
The stability of a given ground state is determined by the grains $n$
where $h_n$ and $j_n$ are comparable, i.e., $\abs{h_n}\sim g\exi\ll1$.
Such sites with very small $h_n$ fields are nothing but the
{\it nucleation sites} which dominate
the low-temperature behaviour of the unhindered model \cite{IV}.
The typical distance $\L(g)$ between two consecutive nucleation sites diverges
as
\beq
\L(g)\sim\frac{1}{g\exi}
\label{lgdef}
\eeq
in the regime of a weak coupling ($g\exi\ll1$).
Thus -- as one might expect -- the closer $g$ is to zero, the less chance
there is that disorder is nucleated.

As long as the size $N$ of the column is smaller than $\L(g)$, or,
equivalently, the coupling constant $g$ is smaller than
$1/(N\exi)$, the unique ground state of the model is the quasiperiodic
ground state of the unhindered model.
For larger columns, a typical ground state can be thought of
as a sequence of independent quasiperiodic patches of mean length $\L(g)$,
pasted together end on end.
For very large sizes ($N\gg\L(g)$), there is
therefore an exponentially large number of such ground states.
The corresponding configurational entropy:
\beq
\S(g)\sim\frac{1}{\L(g)}\sim g\exi
\eeq
vanishes in the weak-coupling regime ($g\exi\ll1$).

Finally, we can provide an integrated description of the ground states
of an irrational $\eps$ and of its rational approximants.
For $g$ small and $\eps$ a fixed irrational, we consider the
sequence of its rational approximants $\eps_k=p_k/q_k$ \cite{IV}.
The periods $p_k+q_k$ of these approximants typically grow exponentially
fast with the approximant order $k$.
Equation (\ref{gcdef}) implies the following.
For the first rational approximants in the series,
whose periods are smaller than $1/(g\exi)$,
the ground states are the same as in the unhindered model.
For all the higher approximants, whose periods exceed this threshold,
ground states are expected to be made of nearly independent patches,
whose characteristic length $\L(g)$ is given by (\ref{lgdef}),
just as for the limiting irrational.
Notice that the period of the approximant that divides these two behaviours
is of the order of $\L(g)\sim1/(g\exi)$ -- there is a reassuring consistency in
this.

\subsection{More details for $\eps=1$}

We now focus on the simplest case, which is obtained when
the shape parameter equals $\eps=1$.
The complex behaviour we obtain
even from this simplest of all cases is a tribute to the inherent
richness of the model.

\subsubsection*{Generic configurations}

Assume that the column size $N$ is even for definiteness.
Consider first a generic configuration.
Equation (\ref{fdef}) simplifies for $\eps=1$ to $f(\s_n)=-\s_n$.
As a consequence, the components of the local fields read
\beq
h_n=-\sum_{m=1}^{n-1}\s_m,\quad
j_n=-\sum_{m=n+1}^N\exi^{m-n}\s_m.
\label{hjsum}
\eeq
The first expression shows that $h_n$ is an integer, whose parity is
opposite to $n$.
Therefore:

\noindent$\bullet$
When the depth $n=2k$ is even, $h_n$ is odd, and thus always non-vanishing.
In the weak-coupling regime ($g\ll1$), we thus obtain
$H_n\approx h_n$, irrespective of $\xii$.

\noindent$\bullet$
When the depth $n=2k-1$ is odd, $h_n$ is even, and may therefore vanish.
When it does, we have $H_n=gj_n$, so that the sign of $H_n$
depends on $\xii$ in general.

In order to understand the complexity which can arise from
this dependence on the interaction length $\xii$, let us focus on
a column of size $N=6$, in the particular configuration $+-+-++$.
We have $\s_3=+1$, $h_3=0$, and $H_3=gj_3=g\exi(1-\exi-\exi^2)$.
The parenthesis is a second-degree polynomial in $\exi$.
It vanishes (in the physical range $0<\exi<1$) when $\exi$ equals
the inverse golden mean
\beq
\phi=\frac{\sqrt{5}-1}{2}\approx0.61803.
\label{phidef}
\eeq
Thus when $0<\exi<\phi$, $H_3=gj_3$ is positive, and
the condition $\s_3=\sign H_3$ is fulfilled.
This condition does not hold for $\phi<\exi<1$.

This example is illustrative of a general property of the model.
For a finite column made of $N$ grains, the statics and dynamics
of the model depend on the relative position of $\exi$
with respect to a finite number of threshold values,
where quantities of interest are discontinuous in general.
These threshold values are given by the following rule:
they are all the roots (in the physical range $0<\exi<1$)
of all the reduced fields $j_n/\exi$ with $n\ge3$ odd,
viewed as polynomials in $\exi$, in all the configurations.
The polynomial $j_n(\exi)/\exi$ has even degree $d=N-n-1$, so that $d\le N-4$.

These threshold values all lie in the range $1/2<\exi<1$.
This fact can be seen as follows.
The expression (\ref{hjsum}) for the local field $j_n$ can be recast as
\beq
j_n=-\exi\s_{n+1}-\sum_{m=n+2}^N\exi^{m-n}\s_m.
\label{jrecast}
\eeq
The sum in the right-hand side is smaller than $\exi^2/(1-\exi)$
in absolute value.
As a consequence, the local field $j_n$ always has the sign of the leading
term, and therefore cannot vanish,
as long as $\exi>\exi^2/(1-\exi)$, i.e., $\exi<1/2$.
The above property is responsible for a strikingly general result:
no dynamical quantities for arbitrary system sizes
depend on $\exi$ for $0<\exi<1/2$, i.e., for $\xii<\xiione$, with
\beq
\xiione=\frac{1}{\ln2}\approx1.44269.
\label{xiionedef}
\eeq
In particular, this explains the plateau
in the data to be presented in Figure~\ref{figv}.

We have generated all the relevant polynomials up to degree $d=12$.
The numbers $A_d$ of physical roots with degree $d$,
and the numbers $B_N$ of threshold values for a column of $N$ grains,
are found to be the following:\footnote{Since some of the polynomials are
reducible, the same root can be generated several times, so that the number
$B_N$ of distinct thresholds may be smaller than
the sum $A_2+A_4+\cdots+A_{N-4}$.}
\beq
\matrix{
A_2=1,\ A_4=5,\ A_6=19,\ A_8=97,\hfill\cr
A_{10}=442,\ A_{12}=1880,\hfill\cr
B_2=0,\ B_4=0,\ B_6=1,\ B_8=6,\ B_{10}=25,\hfill\cr
B_{12}=121,\ B_{14}=563,\ B_{16}=2443.\hfill}
\eeq
We can extract several insights from the above numbers.
Since $B_2=B_4=0$, columns of size $N=2$ and $N=4$
exhibit no dependence on $\xii$ at all.
For $N=6$, we have $B_6=1$:
the unique threshold value is the inverse golden mean $\phi$
introduced in (\ref{phidef}).
Consequently, quantities typically assume two different forms in the intervals
$0<\exi<\phi$ or $\phi<\exi<1$.
For $N=8$, we have $B_8=6$:
quantities typically assume seven different forms in the intervals
demarcated by the six threshold values of $\exi$, and so on.
The interested reader is referred to
Appendix A for the explicit verification of these predictions
for columns of size $N=4$ and $N=6$.

\subsubsection*{Ground states}

We now focus on the ground states in the simple case where $\eps=1$.
They consist of {\it dimerised} configurations,
made of the patterns $P_1=+-\null$ and $P_2=-+\null$.
Assuming again that the column depth $N$ is even for definiteness,
the generic ground state can be described as follows:
\beq
\s_{2k-1}=-\s_{2k}=\eta_k\quad(k=1,\dots,N/2),
\label{gs1}
\eeq
with the dimer variable $\eta_k=+1$ (resp. $\eta_k=-1$)
corresponding to the $P_1$ (resp. $P_2$) pattern.
The boundary condition (\ref{init}) implies $\eta_1=+1$.
The local fields in a ground state read as follows:
\beq
\matrix{
h_{2k-1}=0,\quad h_{2k}=-\eta_k,\hfill\cr
\ds{j_{2k-1}=\exi\eta_k-(1-\exi)\sum_{l=k+1}^{N/2}\exi^{2l-2k}\eta_l},\hfill\cr
\ds{j_{2k}=-(1-\exi)\sum_{l=k+1}^{N/2}\exi^{2l-2k-1}\eta_l}.\hfill}
\label{hjgsdef}
\eeq
There are $2^{N/2}$ ground states in total,
or $2^{N/2-1}$ if (\ref{init}) is taken into account.

The two crystalline (uniform) ground states
\beq
\matrix{
U_+=+-+-+-+-+-\cdots,\cr U_-=-+-+-+-+-+\cdots\hfill}
\label{unidef}
\eeq
play a special role; intuitively, this is because any other ground state has
conflicts between successive dimer pairs, for example in a configuration
such as $+--+$ where the second and third orientations
would be in non-ideal positions.
Such conflicts would of course also be present, albeit more weakly,
between dimer pairs that were more distant from each other along
the column, e.g. $+-\cdots-+$.

This observation can be turned to a quantitative classification
of ground states by means of the {\it pseudo-energy}.
This quantity is defined for an arbitrary configuration as follows:
\beq
\E=-\sum_{n=1}^NH_n\s_n,
\label{edef}
\eeq
i.e.,
\beq
\E=\E_0+g\E_1,
\eeq
with
\beq
\E_0=-\sum_{n=1}^Nh_n\s_n,\quad\E_1=-\sum_{n=1}^Nj_n\s_n.
\label{ecomp}
\eeq
This definition can be motivated as follows.
If the $\s_n$ were independent spins in external fields $H_n$,
(\ref{edef}) would be the corresponding Hamiltonian.
In the present model,
we recall that the local fields $H_n$ depend on the orientations $\s_m$
in a complex and non-symmetric way,
so that the dynamics does not obey detailed balance,
and the statics is not described by any simple Hamiltonian.
The pseudo-energy defined by (\ref{edef}), however,
provides a useful measure of the amount of configurational disorder
in the full interacting system and, in particular,
allows us to classify the ground states.

In the case of a ground state, the first component of the pseudo-energy
reads $\E_0=-N/2$
(irrespective of the ground state), whereas the second one reads
\beq
\E_1=-\frac{N\exi}{2}-(1-\exi)^2\sum_{1\le k<l\le N}\exi^{2l-2k-1}\eta_k\eta_l,
\label{eone}
\eeq
in terms of the dimer variables $\eta_k$.
The first term in the above expression,
\beq
\mean{\E_1}=-\frac{N\exi}{2},
\label{emeandef}
\eeq
is the mean of $\E_1$, in the sense of a uniform average over all the ground
states.
The second term in (\ref{eone}) represents the fluctuation in $\E_1$
from a ground state to another, which typically grows as $N^{1/2}$.

The crystalline states $U_\pm$ introduced in (\ref{unidef}),
respectively corresponding to $\eta_k=+1$ and $\eta_k=-1$ for all $k$,
are the two absolute (global) minima of the pseudo-energy.
We find
\beq
\E_1(U_\pm)=-\frac{N\exi}{1+\exi}+\frac{\exi(1-\exi^N)}{(1+\exi)^2}.
\eeq
It follows that the crystalline ground states are separated from a bulk
of roughly equivalent metastable states by an extensive pseudo-energy gap
\beq
\Delta\E=\mean{\E_1}-\E_1(U_\pm)\approx\frac{N\exi(1-\exi)}{2(1+\exi)}.
\eeq
It is remarkable that our model generates the kind
of (free) energy landscape that is familiar
in theories of glasses \cite{miguel},
and guessed to be valid for grains in the jamming limit \cite{0},
where crystalline states lie well below a band of metastable states.

Finally, the exact threshold coupling $g_c$
for ground-state stability for $\eps=1$ can be evaluated as follows.
Equation (\ref{hjgsdef}) implies that the state $U_+$ is the first to be
destabilised by an increase of the coupling,
and that its weakest link is the second grain orientation $\s_2$.
The corresponding local field reads $H_2=-1+gj_2$,
where $j_2$ assumes its largest possible value
$j_2^\max=\exi(1-\exi^{N-2})/(1+\exi)$.
The threshold coupling constant
at which this first destabilisation takes place therefore reads
$g_c=1/j_2^\max$.
In the limit of an infinite column,
we therefore obtain the following exact expression for the threshold coupling:
\beq
g_c=\frac{1+\exi}{\exi}=\e^{1/\xii}+1.
\label{gc1}
\eeq
The threshold coupling is found to be a decreasing function
of the interaction length $\xii$, blowing up exponentially at small $\xii$,
and reaching a finite asymptotic value 2 in the $\xii\to\infty$ limit.
It is interesting to observe that the general bound (\ref{gcdef}),
i.e., $g_c>\e^{1/\xii}/2$ in the present case ($p=q=1$),
captures the qualitative features of the dependence
of the exact result on $\xii$.

\section{Zero-temperature dynamics}
\setcounter{equation}{0}
\def\theequation{4.\arabic{equation}}

The application of zero-temperature dynamics is the canonical way
of finding the ground states of a model.
In the present model, grains are aligned one at a time with their local fields.
More precisely, grain $n$ is selected at a rate given by (\ref{odef}),
and its orientation variable $\s_n$
is aligned along the local field $H_n$ introduced in (\ref{hdef}),
according to the deterministic rule
\beq
\s_n\to\sign H_n.
\eeq
This rule is well-defined for a non-zero coupling constant $g$,
because the local fields $H_n$ generically do not vanish.

This leads to {\it metastability} in
the following sense: a finite column of $N$ grains in an arbitrary initial
configuration is eventually driven
to an absorbing configuration or {\it attractor},
in a finite {\it jamming time} $T$.
This attractor is necessarily one of the ground states described earlier,
i.e., a configuration where every orientation $\s_n$ is aligned with $H_n$.
For a coupling constant $g$ less than the threshold $g_c$ given in (\ref{gc1}),
the attractors are the $2^{N/2}$ dimerised configurations,
whose number is halved to $2^{N/2-1}$ if
the boundary condition (\ref{init}) is taken into account.
Arbitrary initial conditions can lead to any one
of these metastable configurations being reached;
they are however {\it fragile} in the sense that a slightly different
initial condition or stochastic history generically leads to another
attractor being reached instead.
This fragility of metastable states is one of the characteristics
of granular media \cite{0}.

In what follows, we will focus on two aspects of zero-temperature dynamics:

\noindent$\bullet$ {\it Statistics of the jamming time}.
The jamming time $T$ is the random time the system takes to converge to an
attractor, it being understood that the initial configuration is disordered
and randomly chosen.
The $N$ dependence of both the mean jamming time $\mean{T}$
and of its full probability distribution are of interest.
In this respect, we introduce for further reference the reduced variance
\beq
K_T=\frac{\var{T}}{\mean{T}^2}=\frac{\mean{T^2}}{\mean{T}^2}-1.
\label{kdef}
\eeq

\noindent$\bullet$ {\it Statistics of the attractors}.
The statistics of the attractors reached by stochastic dynamics
is also of special interest,
especially in relation to Edwards' flatness hypothesis \cite{edwards}.
We anticipate non-trivial results only in the glassy phase,
which will be investigated in Section 4.4.

\begin{figure}[!t]
\begin{center}
\includegraphics[angle=90,height=4.5truecm]{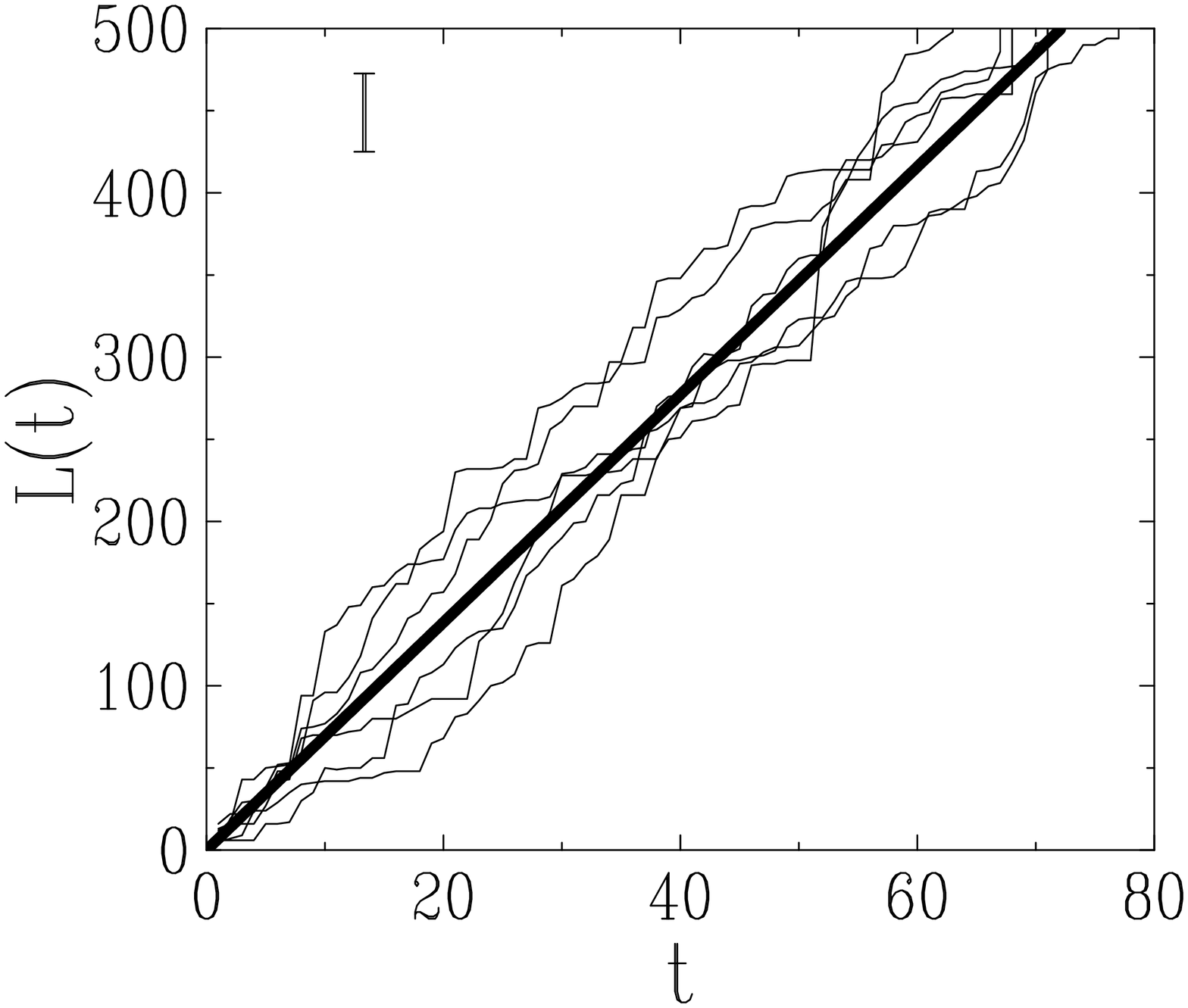}

\includegraphics[angle=90,height=4.5truecm]{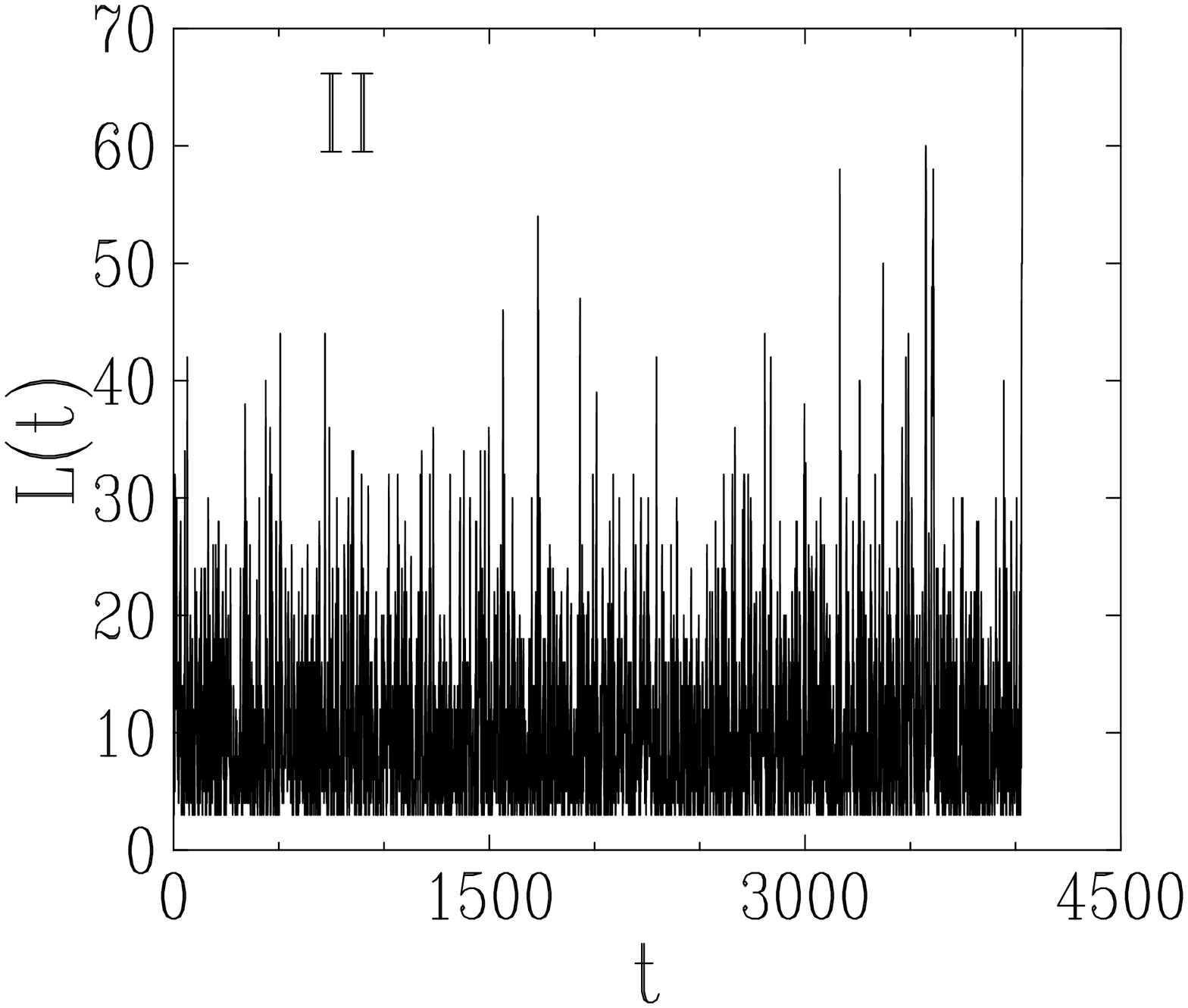}

\includegraphics[angle=90,height=4.5truecm]{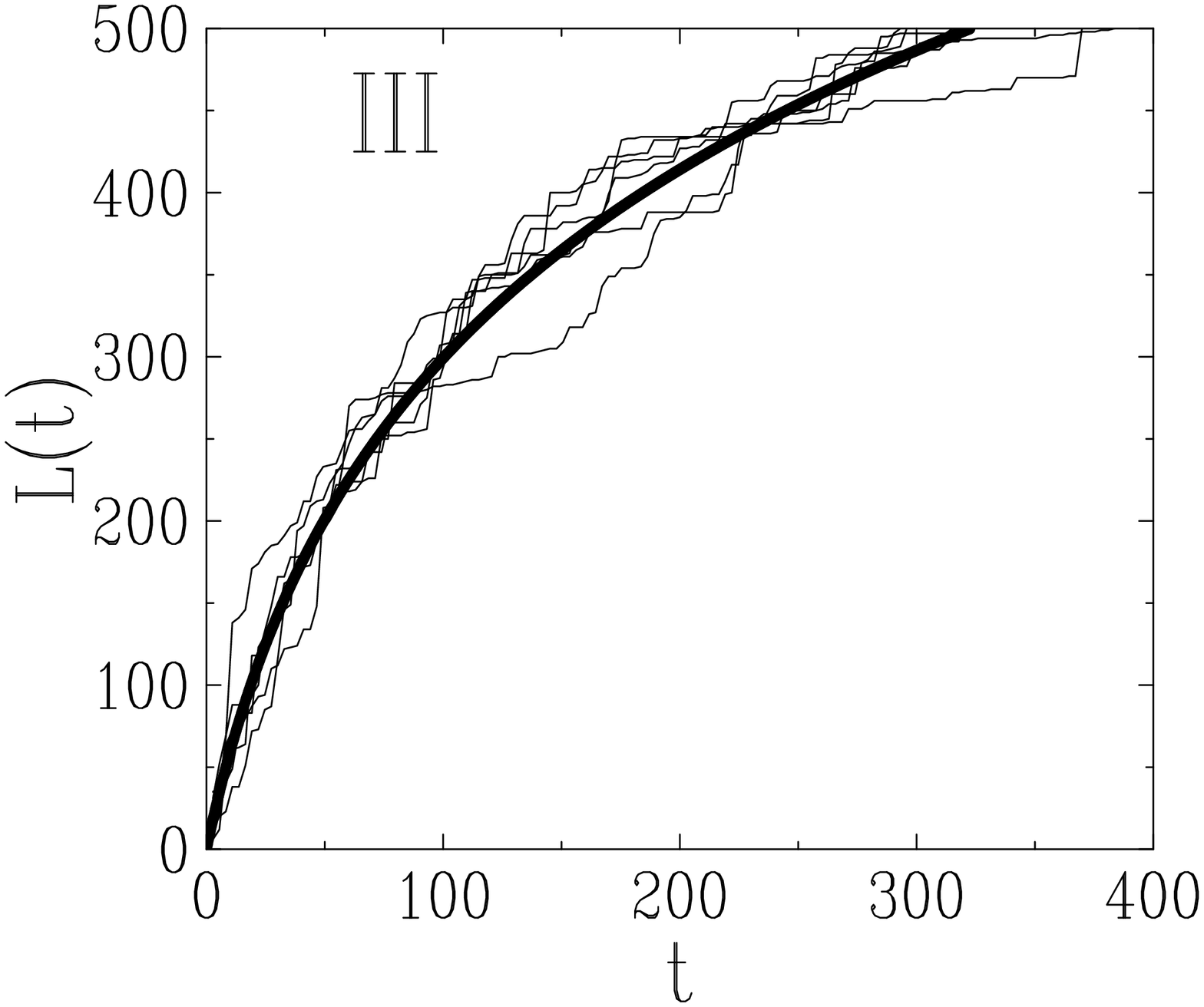}

\includegraphics[angle=90,height=4.5truecm]{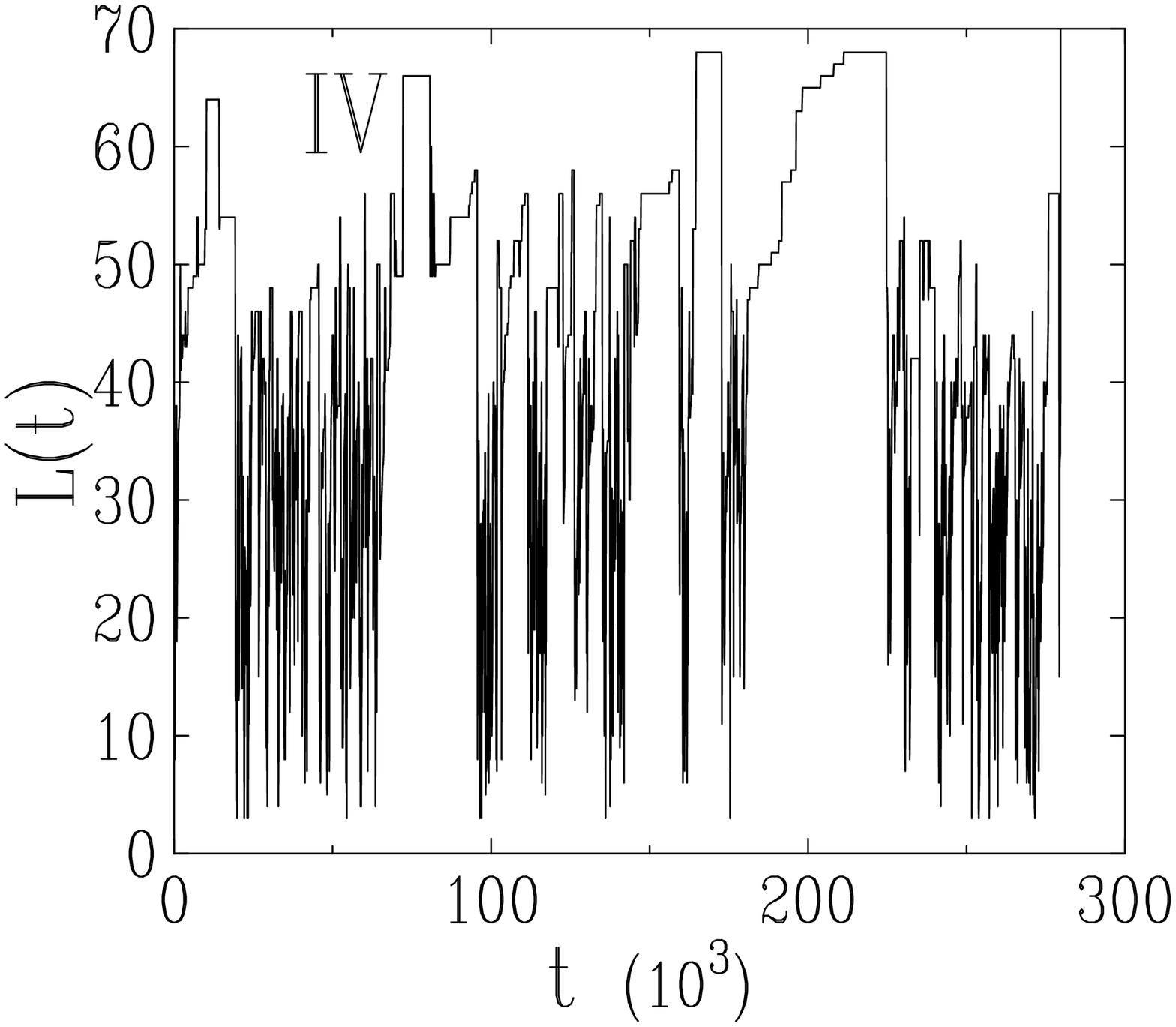}
\end{center}
\caption{\small
Plots of the thickness $L(t)$ of the upper ordered layer of the column
against time $t$, for $g=0.01$, illustrating the four phases.
I: Ballistic phase (several tracks with $\xii=3$, $\xidy=\infty$, $N=500$).
The slope $V=6.95$ of the thick line is taken from Figure~\ref{figv}.
II: Activated phase (one single track with $\xii=50$, $\xidy=\infty$, $N=70$).
III: Logarithmic phase (several tracks with $\xii=3$, $\xidy=200$, $N=500$).
The thick line shows the result (\ref{lxover}) with $V=6.95$.
IV: Glassy phase (one single track with $\xii=50$, $\xidy=7$, $N=70$).}
\label{figabcd}
\end{figure}

\begin{figure}[!t]
\begin{center}
\includegraphics[angle=90,height=3.5truecm]{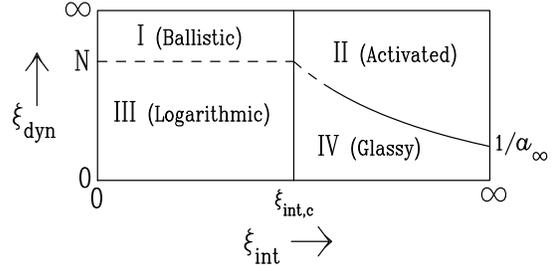}
\caption{\small
Schematic zero-temperature dynamical phase diagram of the model
in the $\xii-\xidy$ plane,
showing the four dynamical phases revealed and investigated below.
The numbers $\xiic$ and $a_\infty$ will be determined later
(see (\ref{xiic}) and (\ref{ainf})).}
\label{figdiag}
\end{center}
\end{figure}

Inspired by previous work \cite{III,IV},
we monitor the various dynamical regimes of the model by means of the thickness
$L(t)$ of the upper ordered layer of the column, defined as the depth
of the uppermost grain which is not aligned with its local field:
\beq
L(t)={\rm inf}\{n\st\s_n\ne\sign H_n\}.
\label{ldef}
\eeq

Figure~\ref{figabcd} shows typical tracks $L(t)$ for four representative
choices of values of $\xii$ and $\xidy$.
Here and throughout the following, we choose for definiteness
the value $g=0.01$ of the coupling constant.
This value is deep in the weak-coupling regime,
where all the results are virtually independent of the precise choice made
for the coupling constant\footnote{For comparison, we recall that the threshold
coupling $g_c$ (see (\ref{gc1})) is always larger than 2.}.
The dynamical behaviour observed turns out to be
very strongly dependent on the lengths $\xii$ and $\xidy$, and
suggests the
existence of four qualitatively different dynamical phases, which
we have named {\it ballistic, logarithmic, activated} and {\it glassy}.
They will be investigated in greater detail in what follows.
The dynamical phase diagram
in the $\xii-\xidy$ plane presented in Figure~\ref{figdiag} shows
already the existence of
genuine phase boundaries (where crossover phenomena become arbitrarily sharp
in the limit of an infinite column) denoted by full lines,
and crossover phenomena (which occur
when $\xidy$ is comparable with the column size $N$) indicated by dashed lines.

\subsection{\bf Phase I (Ballistic)}

This phase is illustrated by panel I of Figure~\ref{figabcd} -- it
corresponds to small $\xii$ and large $\xidy$.
Physically this implies that one is looking at layers
near the free surface (large $\xidy$) of a column where grains feel
correlations from below only weakly (small $\xii$).
In other words, we are looking at the `top' of a granular column.

The thickness $L(t)$ is observed to grow on average linearly with time:
\beq
\mean{L(t)}\approx Vt.
\label{vt}
\eeq
The (relatively small) fluctuations between different tracks
correspond to different stochastic histories.
Equation (\ref{vt}) shows that an ordered layer propagates
ballistically down the column with velocity $V$, a phenomenon which
was already encountered in the unhindered model \cite{III,IV}.
When $L(t)$ becomes equal to the column depth $N$,
an attractor is reached and the dynamics stops, so that
\beq
T\approx\frac{N}{V},
\label{fbal}
\eeq
again up to relatively negligible fluctuations.

\begin{figure}[!t]
\begin{center}
\includegraphics[angle=90,height=5truecm]{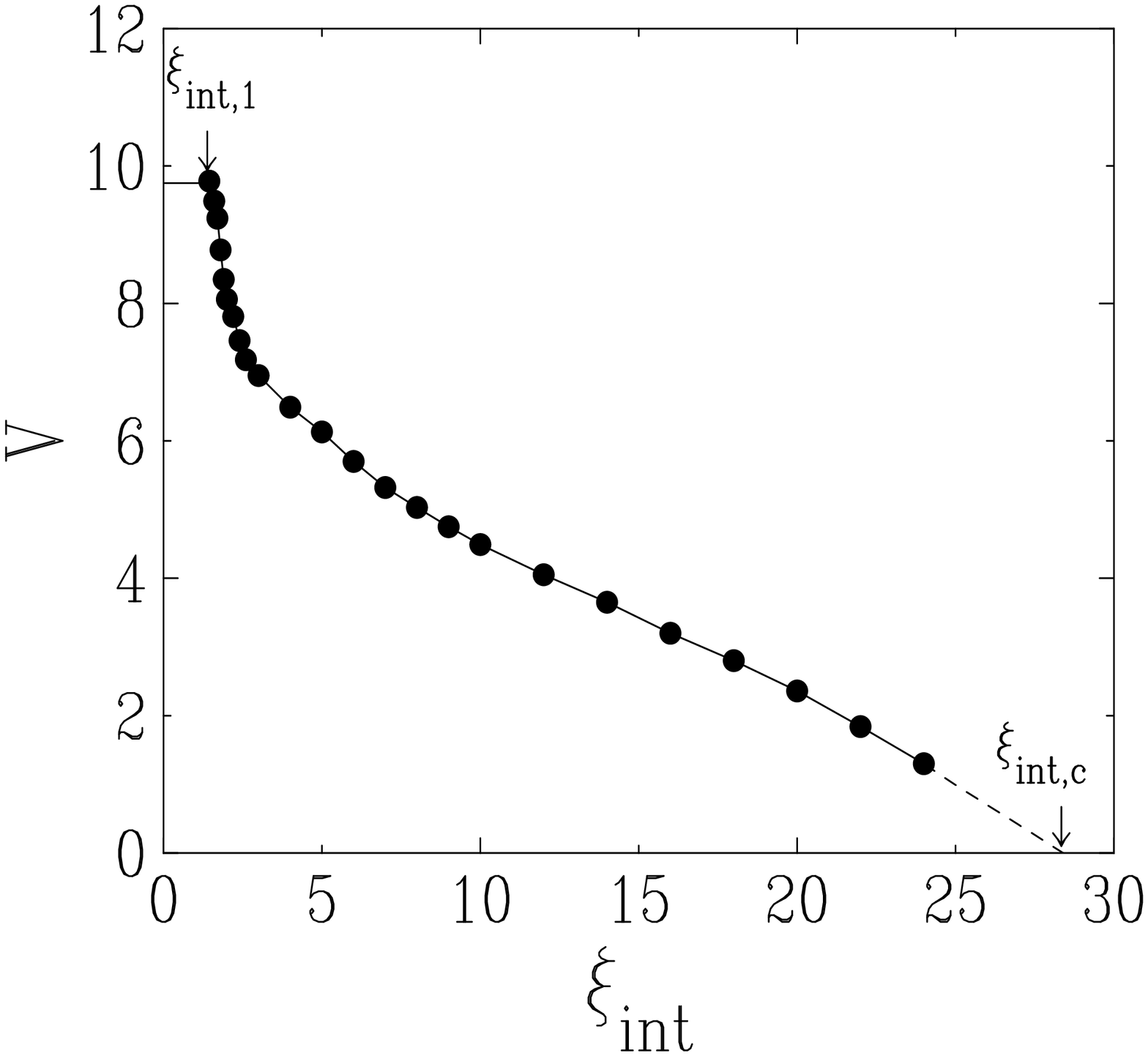}
\caption{\small
Plot of the ballistic velocity $V$ against $\xii$ for $\xidy=\infty$.
Arrows indicate the value $\xiione$ (see (\ref{xiionedef}))
below which $V$ is constant and equal to $V_0$ given in (\ref{vzero}),
and the critical point $\xiic$ (see (\ref{xiic}))
at which $V$ vanishes linearly.}
\label{figv}
\end{center}
\end{figure}

Figure~\ref{figv} shows numerical values of the velocity $V$,
obtained by averaging $L(t)$ over many independent initial configurations
and histories (at least $10^4$ per point).
The first interesting feature is a plateau region observed for
$\xii$ smaller than the threshold value $\xiione$
determined in (\ref{xiionedef}).
As predicted in Section 3.3, the velocity $V$ is found to be constant
over this region, and equal to its value in the $\xii\to0$ limit:
\beq
V_0\approx9.75.
\label{vzero}
\eeq
As $\xii$ increases, the effects of frustration begin to kick in more and more
via the back-reaction $j_n$; the resulting inefficiency of the zero-temperature
dynamics causes the velocity to decrease progressively with $\xii$.

It seems strange that $V_0$ is such a large number, especially given that
it leads to the apparition of other large dimensionless numbers,
such as $\xiic\approx28.4$ (see (\ref{xiic}))
or $D_c\approx37$ (see (\ref{critd})).
Fortunately, we can provide a simple explanation for the high value of $V_0$,
that the naturalness principle \cite{thooft} would demand.
We recall (see the lines following (\ref{jrecast}))
that the component $j_n$ of the local field
always takes the sign of $-\s_{n+1}$ for $\xii<\xiione$.
Assume that the uppermost $2k$ grains of the column are already dimerised.
So far as zero-temperature dynamics is concerned,
the next two orientations $\s_{2k+1}$ and $\s_{2k+2}$,
are entirely decoupled from the rest of the column.
Indeed, $h_{2k+1}=0$, so that $H_{2k+1}=gj_{2k+1}$ has the sign of $-\s_{2k+2}$,
whereas $H_{2k+2}\approx h_{2k+2}=-\s_{2k+1}$.
If we make the simple assumption that the four configurations
of these two orientations are equally probable,
the mean time it takes to go to one of the two possible attractors
can be shown to be $1/4$.
This newly formed dimer has a spatial size 2,
and the corresponding front velocity is $2/(1/4)=8$.
The actual velocity $V_0$ of (\ref{vzero}) is only about 21\% above this
simple-minded estimate.\footnote{We mention for comparison that in the
analogous case in the model of
\cite{III,IV}, that of an irrational $\eps\to1^\pm$
in the immediate neighbourhood of $\eps=1$, a similar line of reasoning yields
the value 2 for the velocity, whereas the measured velocity
$V_{1^\pm}\approx2.38$ is about 19\% above that estimate.}.

The data of Figure~\ref{figv} also show that
$V$ vanishes linearly as the borderline between Phases I and II
is approached, i.e., as $\xii\to\xiic^-$
This behaviour fits in a natural way within
the effective description put forward in Section 4.2
in terms of biased Brownian motion (see (\ref{critv})).

\subsection{\bf Phase II (Activated)}

This phase of relatively large $\xii$ and $\xidy$
is illustrated by panel II of Figure~\ref{figabcd}.
One is still considering grains that are relatively free to move,
as in the top layers of a column, but now grains are increasingly constrained
as a result of grain orientations below them.

There is, typically, only very weak order in the column
before it happens to jam: this is exemplified by a mean thickness
$\mean {L(t)}$ which is quite small compared to the column depth $N$.
Also, $L(t)$ exhibits wild fluctuations around its mean,
which look stationary over the very long time it takes for the system to jam.
After sporadic excursions to larger values, the layer thickness
suddenly jumps to $L(t)=N$, so that an attractor is reached.

This phenomenology is typical of an activated phenomenon.
We therefore expect that:

\noindent$\bullet$
The statistics of the jamming time $T$ should be approximately given
by an exponential distribution:
\beq
\rho(T)=\frac{1}{\mean{T}}\exp\left(-\frac{T}{\mean{T}}\right),
\label{expo}
\eeq
characterised by a single scale $\mean{T}$,
with unit reduced variance ($K_T=1$).

\noindent$\bullet$
The mean jamming time should grow exponentially with the column size:
\beq
\mean{T}\sim\exp(a(\xii)N),
\label{fact}
\eeq
at least for very large $N$,
where $a(\xii)$ is the reduced activation energy per grain,
i.e., the height of the entropic barrier the system has to cross
to reach the ground state.

Huge finite-size effects rule out an accurate numerical exploration
of the activated phase for generic $\xii$ and $\xidy$.
In the following, setting $\xidy\to\infty$,
we examine two limits of particular interest.
We explore first the crossover between Phases I (ballistic)
and II (activated), as $\xii$ approaches the critical value $\xiic$.
Next, we examine the regime of deep activation, when the effect
of upward interactions is maximal ($\xii\gg N$).

\subsubsection*{Crossover between Phases I and II}

In order to understand the crossover between Phases I and II,
we invoke the following picture of the behaviour of the thickness $L(t)$
of the ordered layer.
In either of the two phases, it starts from the surface
and eventually propagates to the base, when an attractor is reached.
In the purely ballistic case ($\xii$ very small),
the layer essentially shoots down to form an attractor.
The effect of increasing $\xii$ is to `admit impediments' to this pure flow,
to cause $L(t)$ to fluctuate (diffuse) increasingly before the whole column
reaches an attractor.
The uniform effect of grains above any given grain
and the back-reaction of the grains below it are responsible for the
frustration that is increasingly encountered in the search for an attractor
as $\xii$ increases.
The value of $\xii$ at which both interactions balance out
is the critical point $\xiic$, at which the velocity $V$ vanishes,
so that the dynamics is purely diffusive.

\begin{figure}[!t]
\begin{center}
\includegraphics[angle=90,height=5truecm]{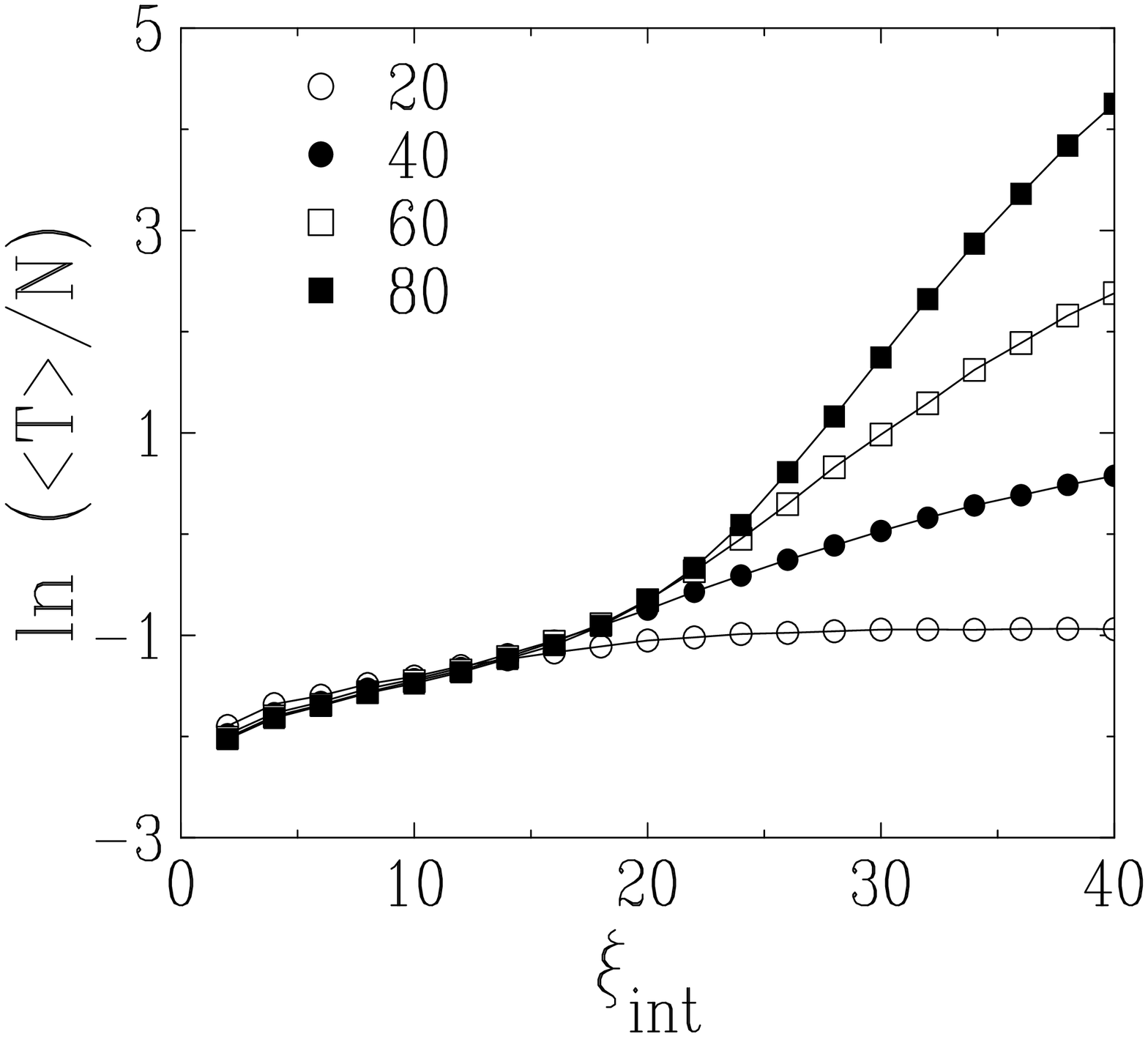}

\includegraphics[angle=90,height=5truecm]{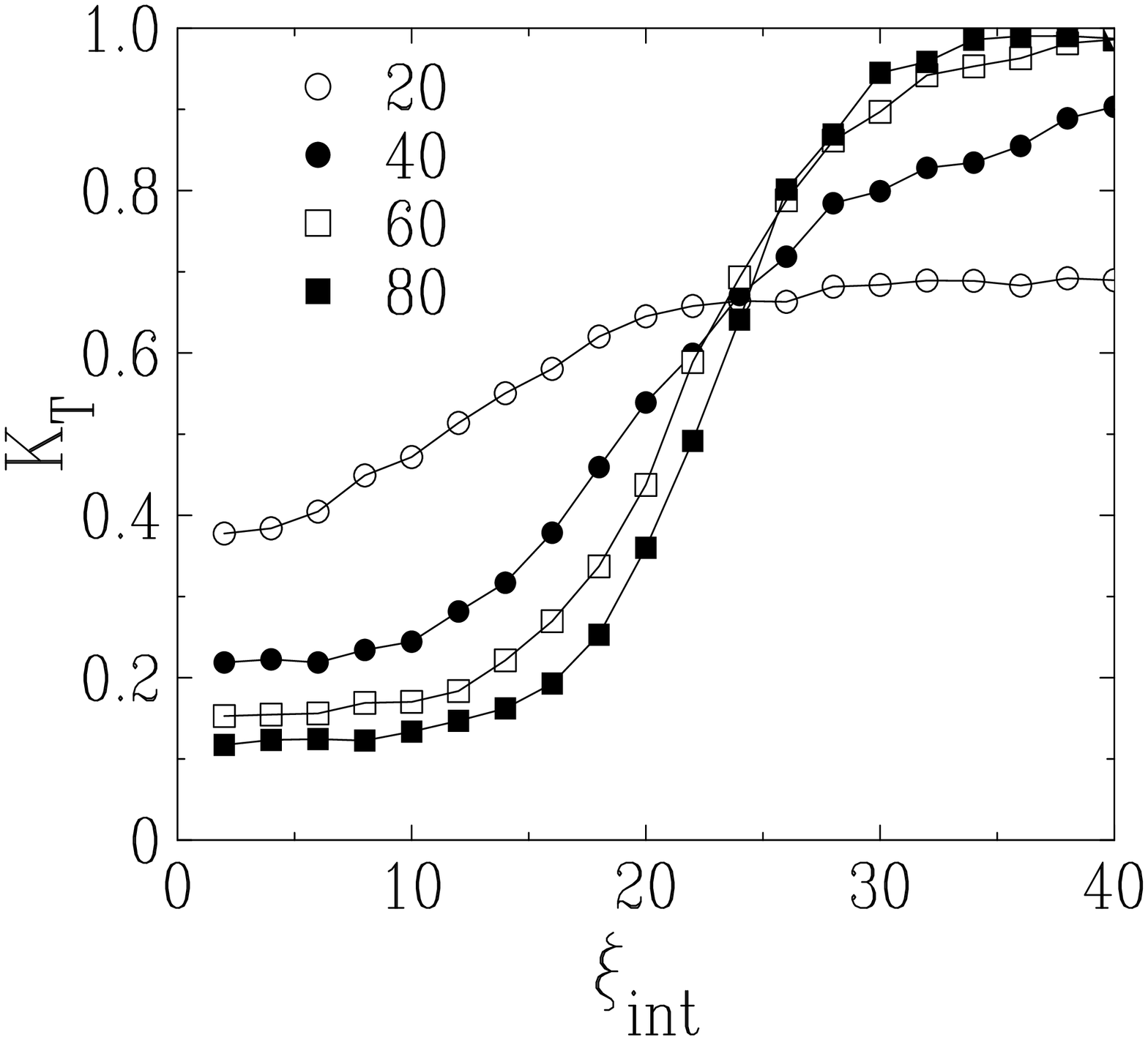}
\caption{\small
Top: logarithmic plot of the ratio $\mean{T}/N$ against $\xii$,
for a dynamical length $\xidy=\infty$, and variable column size $N$.
Bottom: plot of the reduced variance $K_T$ against $\xii$,
for the same parameters.}
\label{figtk}
\end{center}
\end{figure}

The above intuitive picture is corroborated by Figure~\ref{figtk},
showing a logarithmic plot of the ratio $\mean{T}/N$
and a plot of the reduced variance $K_T$ against $\xii$,
for several values of the column size $N$.
There is evidence of a continuous phase transition
between a ballistic phase for $\xii<\xiic$ and an activated phase
for $\xii>\xiic$.
We notice from the top panel that $\mean{T}/N$ is roughly independent of $N$
in the ballistic phase, whereas it grows fast with $N$ in the activated phase.
On the other hand, the plots of $K_T$ approximately cross at a critical value
$K_T\sim0.7$ for $\xii=\xiic\sim26$.

This picture can be turned into the following effective model.
We treat the thickness $L(t)$ as a collective coordinate,
and model its dynamics by a biased Brownian motion on an interval,
with velocity $V$ and diffusion constant $D$.
The motion starts at time $t=0$ at an initial point $L(0)$
very near the free surface of the column,
which is considered as a reflecting boundary.
It ends at the random hitting time $t=T$
when $L(t)$ visits the base of the column, i.e., $L(t)=N$, for the first time.
Accordingly, the base is considered as an absorbing boundary.
This effective model is analysed in detail in Appendix B.

\begin{figure}[!t]
\begin{center}
\includegraphics[angle=90,height=5truecm]{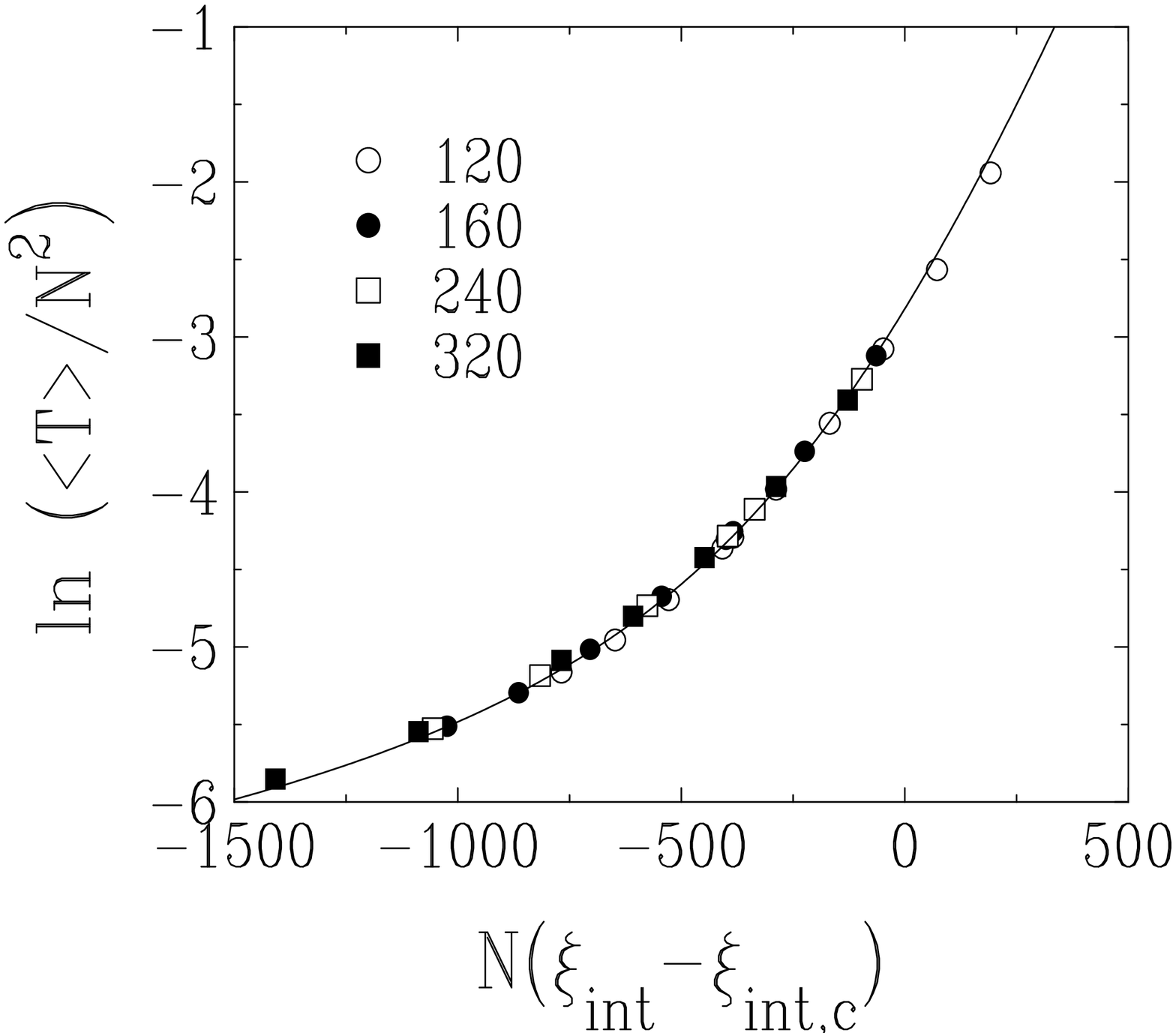}

\includegraphics[angle=90,height=5truecm]{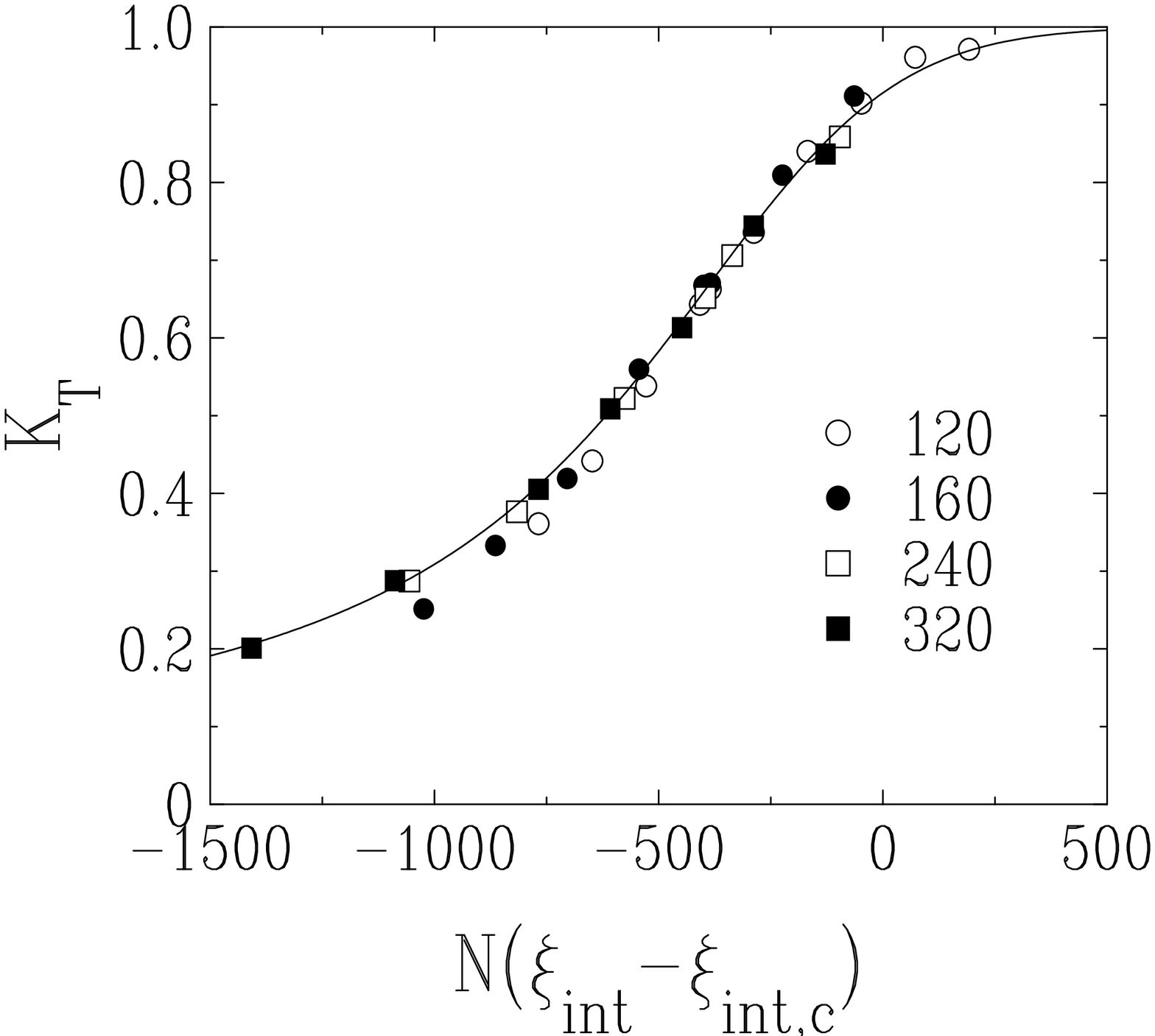}
\caption{\small
Top: logarithmic plot of $\mean{T}/N^2$ against $N(\xii-\xiic)$.
for a dynamical length $\xidy=\infty$, $\xiic=28.4$, and variable $N$.
Bottom: plot of the reduced variance $K_T$ against $X$,
for the same parameters.
Full lines: plots of the analytical
results (\ref{fssf}) and (\ref{fssg}) for the effective model,
rescaled according to (\ref{rescaldef}).}
\label{figfg}
\end{center}
\end{figure}

Figure~\ref{figfg} shows that both the mean jamming time $\mean{T}$
and its reduced variance $K_T$ obey finite-size scaling laws of the form
\beq
\mean{T}\approx N^2\,F(X),\quad K_T\approx G(X),
\label{fssdef}
\eeq
with
\beq
X=N(\xii-\xiic).
\label{Xdef}
\eeq
The best data collapse is obtained for
\beq
\xiic\approx28.4.
\label{xiic}
\eeq

Furthermore, the data are observed to be in accurate agreement
with the finite-size scaling results (\ref{fssf}) and (\ref{fssg}),
derived analytically in Appendix B for the effective model.
The excellent quality of the accord suggests that the effective model predicts
the exact finite-size scaling functions of the model.
The best agreement, shown as full lines in Figure~\ref{figfg},
is obtained with the following identification:
\beq
X\approx-115(z+3.4),\quad\ln(T_0/N^2)\approx-4.3
\label{rescaldef}
\eeq
between, on the empirical side,
the scaling variable $X$ of the column model introduced in (\ref{Xdef}),
and, on the theoretical side, the scaling variable
\beq
z=\frac{VN}{D}
\eeq
of the effective model, introduced in (\ref{zeddef}),
and the diffusive time scale
\beq
T_0=\frac{N^2}{2D},
\eeq
introduced in (\ref{btc}).
This last relation, together with the second equation of (\ref{rescaldef}),
enable us to predict
the critical value $D_c$ of the diffusion constant for $\xii=\xiic$.
We thus obtain $D_c\approx\e^{4.3}/2$, i.e.,
\beq
D_c\approx37.
\label{critd}
\eeq
The reduced variance of the jamming time at the critical point
is predicted in (\ref{bkc}) to be a universal number:
\beq
K_T=\frac{2}{3}.
\eeq
This prediction is again in good agreement with the apparent crossing point
of the data of the lower panel of Figure~\ref{figtk}.

Finally, we compare the predictions of the effective model in
the ballistic and activated phases with the above results.

\noindent$\bullet$
{\it Toward the ballistic phase ($\xii<\xiic$, i.e., $V>0$)}.
In the ballistic phase, the prediction (\ref{btbal}) for the mean jamming time:
\beq
\mean{T}\approx\frac{N}{V}-\frac{D}{V^2}
\eeq
exhibits the observed ballistic behaviour (\ref{fbal}),
up to a finite negative correction due to diffusion.

The fluctuations of the jamming time around its mean are predicted
to be Gaussian, with a reduced variance given by (\ref{bkbal}):
\beq
K_T\approx\frac{2D}{VN}.
\eeq
This fall-off as $1/N$ agrees with the observation made above
that fluctuations become relatively negligible for large columns.

The critical regime of the ballistic phase corresponds
to the regime $X\to-\infty$, i.e., $z\to+\infty$,
in the finite-size scaling laws (\ref{fssdef}).
Equations (\ref{fssf}) and (\ref{rescaldef}) imply that
the scaling function $F(X)$ falls off as
$F(X)\approx A_F/\abs{X}$, with $A_F\approx115/D\approx3.1$.
This estimate implies in turn
that the velocity vanishes linearly as $\xii\to\xiic^-$,
as $V\approx(\xiic-\xii)/A_F$, i.e.,
\beq
V\approx0.32(\xiic-\xii).
\label{critv}
\eeq
The numerical value of the prefactor is in good agreement with the slope
of the extrapolation curve shown as a dashed line in Figure~\ref{figv}.

\noindent$\bullet$
{\it Toward the activated phase ($\xii>\xiic$, i.e., $V<0$)}.
In the activated phase, the prediction (\ref{bta}) for the mean jamming time:
\beq
\mean{T}\approx\frac{D}{V^2}\;\e^{\abs{V}N/D}
\eeq
grows exponentially with $N$, as anticipated in (\ref{fact}).
The corresponding activation energy per unit length,
\beq
a=\frac{\abs{V}}{D},
\eeq
is essentially given by the negative of the velocity.
The reduced variance of the jamming time predicted by (\ref{bk}):
\beq
K_T\approx1-\frac{2(VN+3D)}{D}\;\e^{-\abs{V}N/D}
\eeq
converges exponentially fast to its limiting value unity,
characteristic of an exponential distribution.

The critical regime of the activated phase corresponds
to the regime $X\to+\infty$, i.e., $z\to-\infty$,
in the finite-scaling laws (\ref{fssdef}).
The effective model predicts that the scaling function $G$
has an exponential convergence toward $G(+\infty)=1$,
whereas the scaling function $F$ grows exponentially as
$F\sim\exp(-z)\sim\exp(B_FX)$ with $B_F\approx1/115\approx0.0087$.
These results corroborate our expectations,
including (\ref{expo}) and (\ref{fact}).
They also imply that the activation energy per grain
vanishes linearly as $\xii\to\xiic^+$,
as $a(\xii)\approx B_F(\xii-\xiic)$, i.e.,
\beq
a(\xii)\approx0.0087(\xii-\xiic).
\label{crita}
\eeq

Before leaving this topic, we emphasise that the simple picture of a
Brownian particle whose velocity $V$ changes sign at $\xiic$
appears to explain all our observations on this crossover.

\subsubsection*{Limiting behaviour for $\xii\gg N$}

We now look at the slowest possible dynamics in the activated phase;
this will clearly occur when the column is at its most correlated,
where $\xii$ is much larger than $N$, still keeping $\xidy=\infty$ for
simplicity.
We are thus led to investigate the doubly singular limit where
$\xidy=\xii=\infty$.
The only free parameter is then the column size $N$.

\begin{figure}[!t]
\begin{center}
\includegraphics[angle=90,height=5truecm]{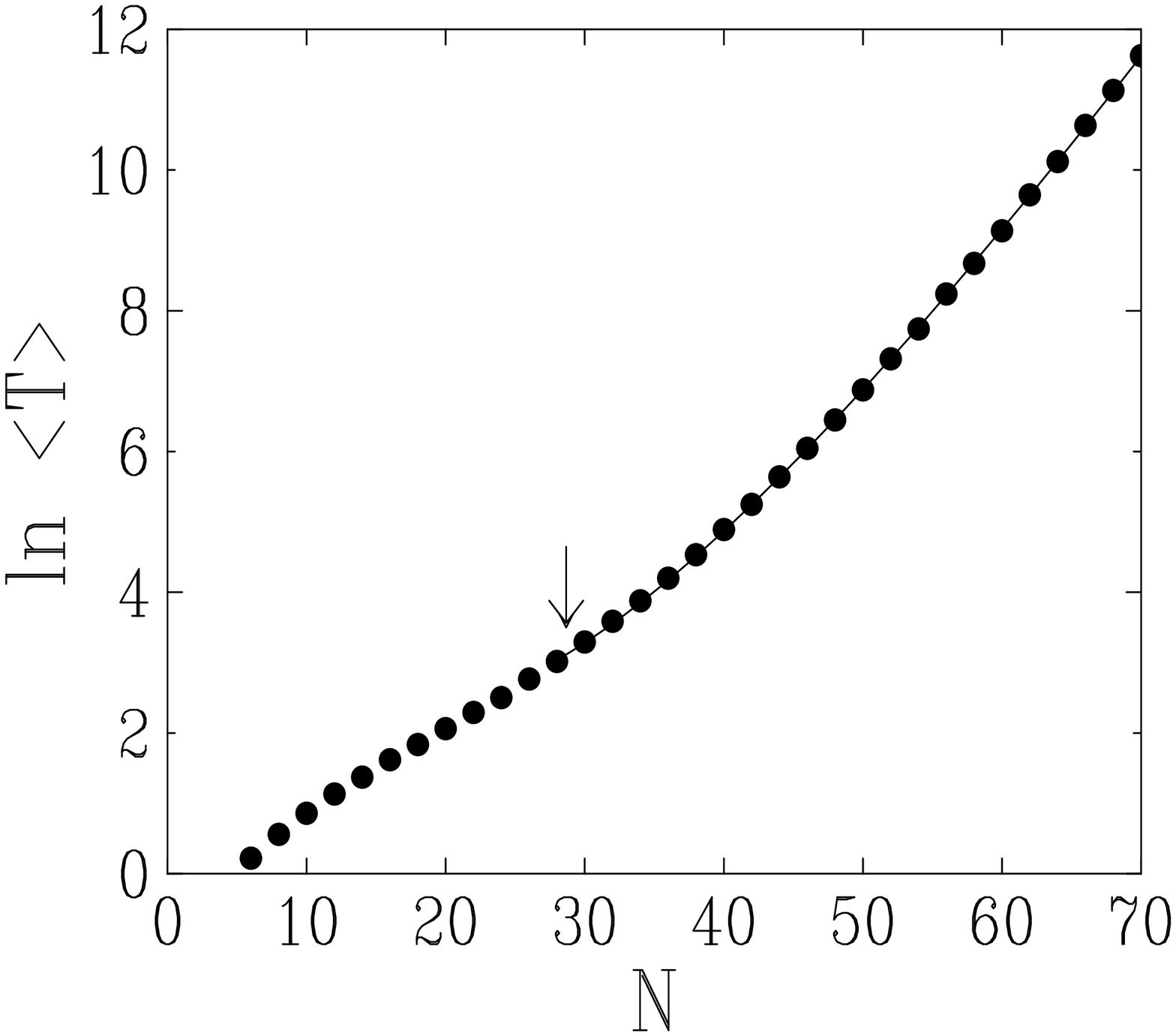}

\includegraphics[angle=90,height=5truecm]{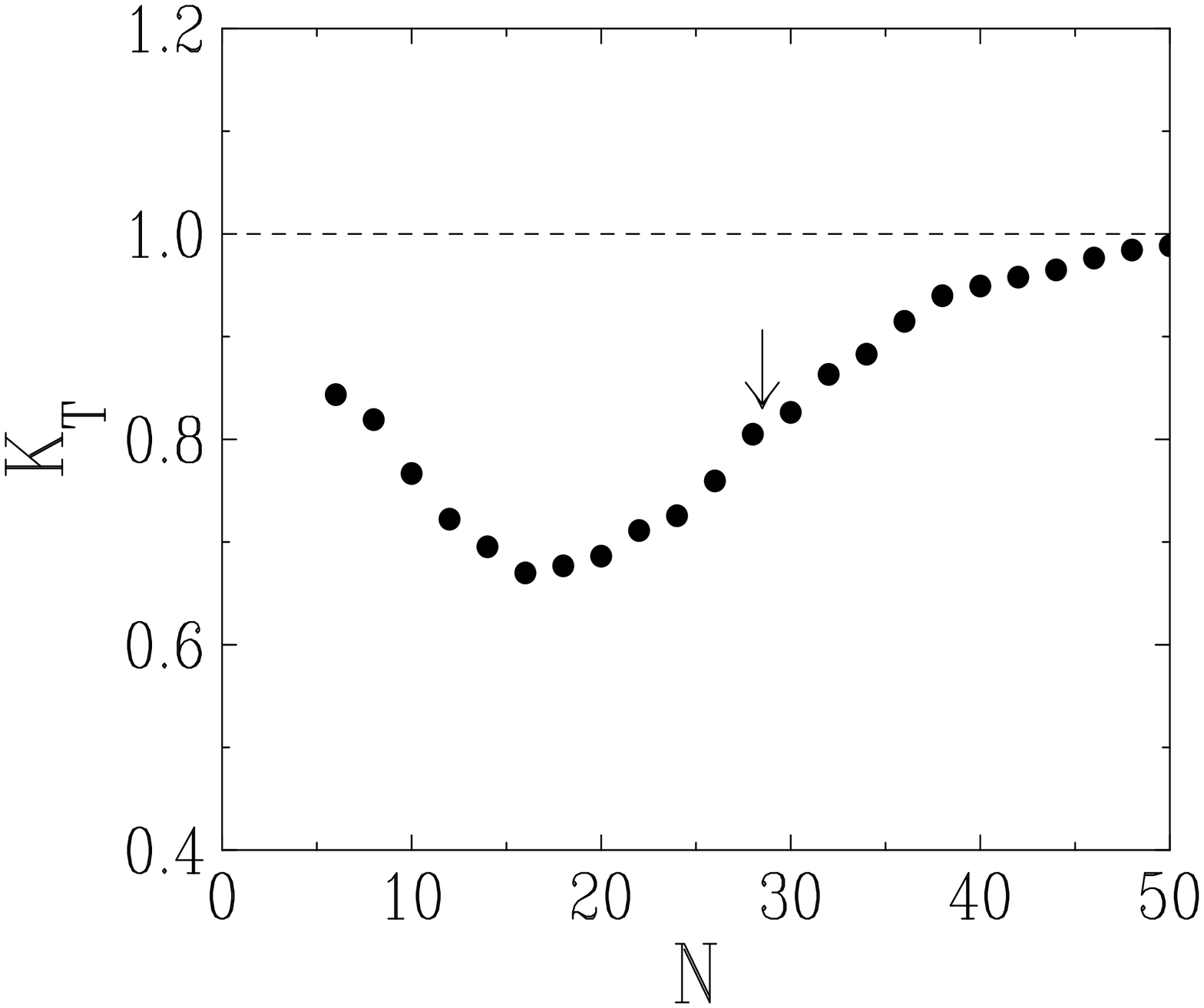}
\caption{\small
Top: logarithmic plot of the mean jamming time $\mean{T}$ against $N$,
for $\xii=\xidy=\infty$.
Full line (hardly visible):
fit $\mean{T}=N(\ln 2)/2-6.515\ln N+15.04$ to the data for $N>\xiic$.
Bottom: plot of the reduced variance $K_T$ against $N$,
for the same parameters.
Arrows show the crossover scale $N=\xiic$ (see (\ref{xiic})).}
\label{figlntkt}
\end{center}
\end{figure}

Figure~\ref{figlntkt} shows plots of the mean jamming time $\mean{T}$
and of its reduced variance $K_T$ against $N$.
When the column size $N$ is rather small,
the system still behaves more or less ballistically;
this fast dynamics leads to the nearly linear growth of $\mean{T}$ with $N$,
and the concomitant decrease of the variance $K_T$ as a function of $N$.
When $N$ is large enough, the system is fully activated, so that the jamming
time grows exponentially with $N$, while the variance increases, rapidly
converging to its asymptotic value $K_T=1$.
The crossover between these behaviours occurs when
the column size $N$ is of the order of $\xiic$ (shown as arrows in both plots).

Finally, we focus on the activation energy $a(\xii)$ defined in (\ref{fact}).
When $\xii\to\infty$, the column is at its most activated; hitting an
attractor is then a totally random process.
The jamming time is therefore expected to be simply given by the ratio between
the number $\Omega_0=2^N$ of disordered initial configurations
and the number $\Omega_\infty=2^{N/2}$ of possible attractors:
\beq
\mean{T}\sim\frac{\Omega_0}{\Omega_\infty}\sim 2^{N/2}.
\label{tentex}
\eeq
A similar purely entropic result is shown in Appendix C to hold
within a toy model of a Markovian dynamics on an assembly
of independent two-level systems.
The result (\ref{tentex}) implies that $a(\xii)$ saturates to the value
\beq
a_\infty=\frac{\ln 2}{2}\approx0.34657.
\label{ainf}
\eeq
This limiting value of the activation energy has been incorporated into
the fit presented in the upper panel of Figure~\ref{figlntkt}.
The good quality of the fit can be viewed as corroborating the result
(\ref{ainf}), in spite of large correction terms.

To sum up, the activation energy $a(\xii)$ is expected to increase
monotonically with $\xii$ all over the activated phase,
and to interpolate smoothly between the linear growth (\ref{crita})
as $\xii\to\xiic^+$ (where the given numerical value of the prefactor only
holds in the $\xidy\to\infty$ limit)
and the purely entropic limiting value (\ref{ainf}) as $\xii\to\infty$.

\subsection{\bf Phase III (Logarithmic)}

This phase of relatively small $\xii$ and $\xidy$
is illustrated by panel III of Figure~\ref{figabcd}.
The thickness $L(t)$ of the ordered layer follows a well-defined master curve,
growing slower than linearly with time,
again with relatively small fluctuations between different tracks.

\begin{figure}[!t]
\begin{center}
\includegraphics[angle=90,height=5truecm]{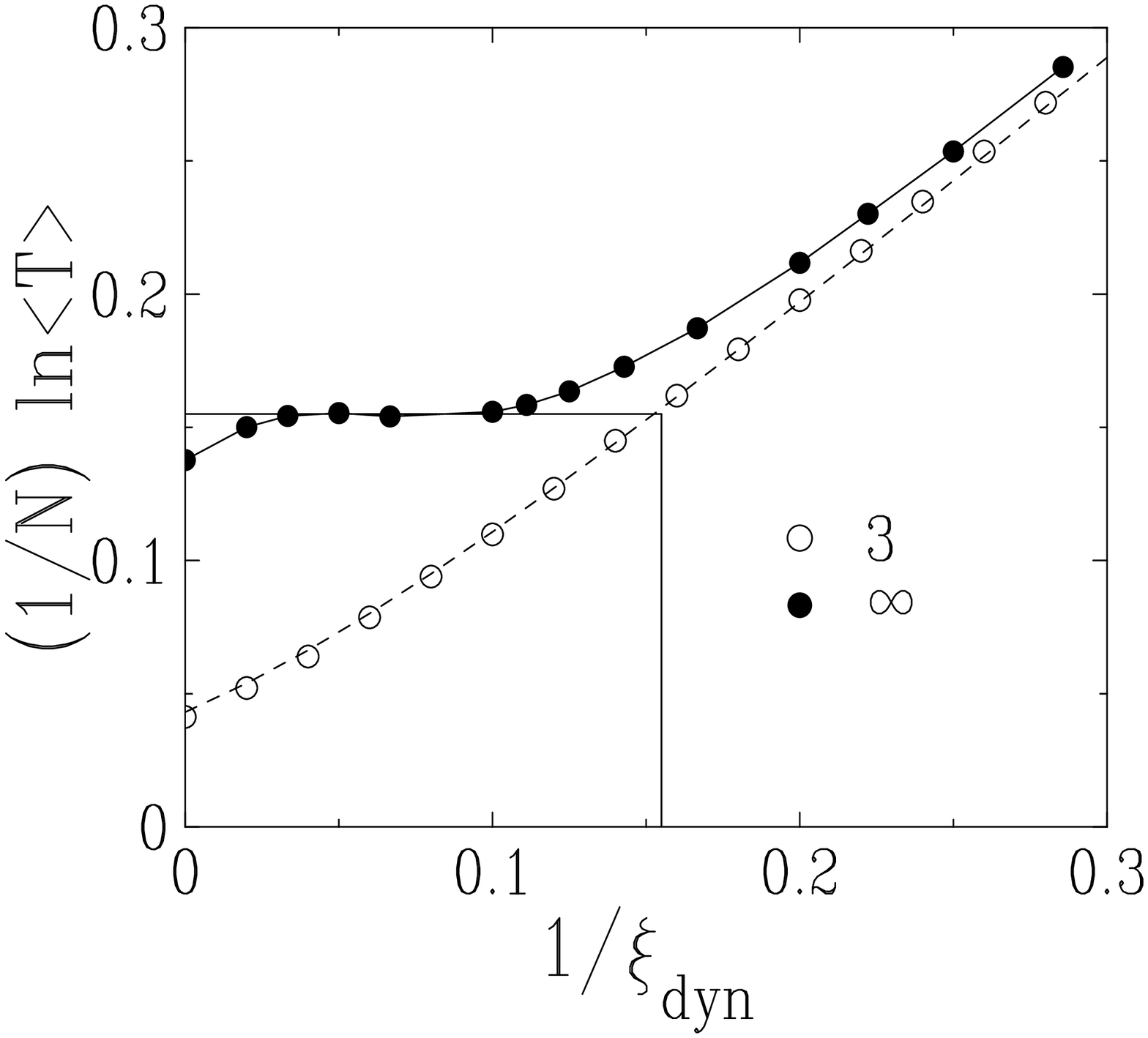}
\caption{\small
Plot of $(1/N)\ln\mean{T}$ against $1/\xidy$ for $N=50$.
Empty symbols ($\xii=3$) demonstrate the crossover between
Phases I (ballistic) and III (logarithmic).
Dashed line: prediction (\ref{xover}), with $V=6.95$.
Full symbols ($\xii=\infty$) corroborate
the crossover (\ref{sharp}) between Phases II (activated) and IV (glassy),
in spite of large finite-size effects.
Full straight lines: crossover at $1/\xidy\approx a_{\rm eff}\approx0.155$
(see text).}
\label{figtim}
\end{center}
\end{figure}

Already encountered in the unhindered model \cite{III,IV},
this phenomenon can be explained as follows.
Equation (\ref{vt}) shows that the application of zero-temperature dynamics
causes order to propagate ballistically,
for $\xii<\xiic$ and $\xidy$ much larger than $N$.
When $\xidy$ becomes comparable with $N$, however, grains
move progressively slowly according to their depth,
with local frequencies that scale as (\ref{odef}).
Writing the differential equation
\beq
\frac{\d L}{\d t}\approx V\,\exp\left(-\frac{L}{\xidy}\right),
\eeq
we find that the thickness grows according to
\beq
L(t)\approx\xidy\ln\left(1+\frac{Vt}{\xidy}\right).
\label{lxover}
\eeq
and the mean jamming time for a column of size $N$ reads
\beq
\mean{T}\approx\frac{\xidy}{V}\left(\exp\left(\frac{N}{\xidy}\right)-1\right).
\label{xover}
\eeq
This result holds all over the left of the phase diagram
of Figure~\ref{figdiag} (Phases I and III and the crossover between them).
It is confirmed quantitatively
by the data shown (empty symbols) in Figure~\ref{figtim}.

The laws (\ref{vt}) and (\ref{fbal}) are recovered for $\xidy\gg N$,
i.e., in the ballistic phase.
In the logarithmic phase, when $\xidy\ll N$,
the width of the ordered layer is predicted to grow logarithmically:
\beq
L(t)\approx\xidy\ln\frac{Vt}{\xidy},
\eeq
so that the mean jamming diverges exponentially fast with the column size $N$:
\beq
\mean{T}\approx\frac{\xidy}{V}\,\exp\left(\frac{N}{\xidy}\right).
\eeq

\subsection{\bf Phase IV (Glassy)}

The glassy phase is found when $\xii$ is large and $\xidy$ is small;
this is by far the richest and most novel phase of this model.
The signal for $L(t)$, illustrated in panel IV of Figure~\ref{figabcd},
is neither nearly deterministic (as in the ballistic and logarithmic phases)
nor totally random (as in the activated phase).
The glassy phase corresponds to the `bottom' of a long column,
where grain reorientations are at their most hindered; grains in
this region are weighed down
by those above them and, additionally, feel to the fullest extent the
effect of the orientational
frustration between upper and lower grains.

\begin{figure}[!t]
\begin{center}
\includegraphics[angle=90,height=4.5truecm]{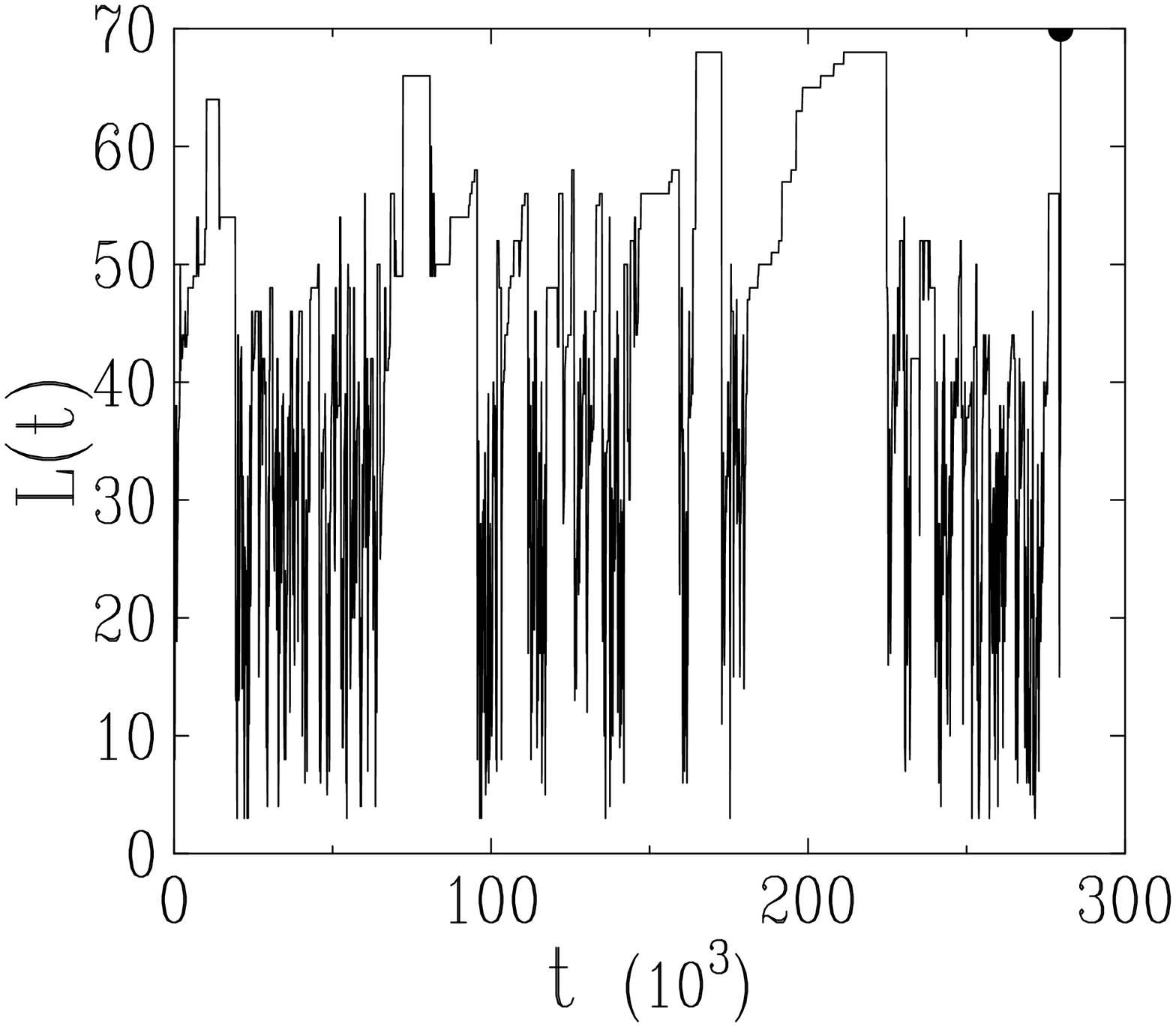}

\includegraphics[angle=90,height=4.5truecm]{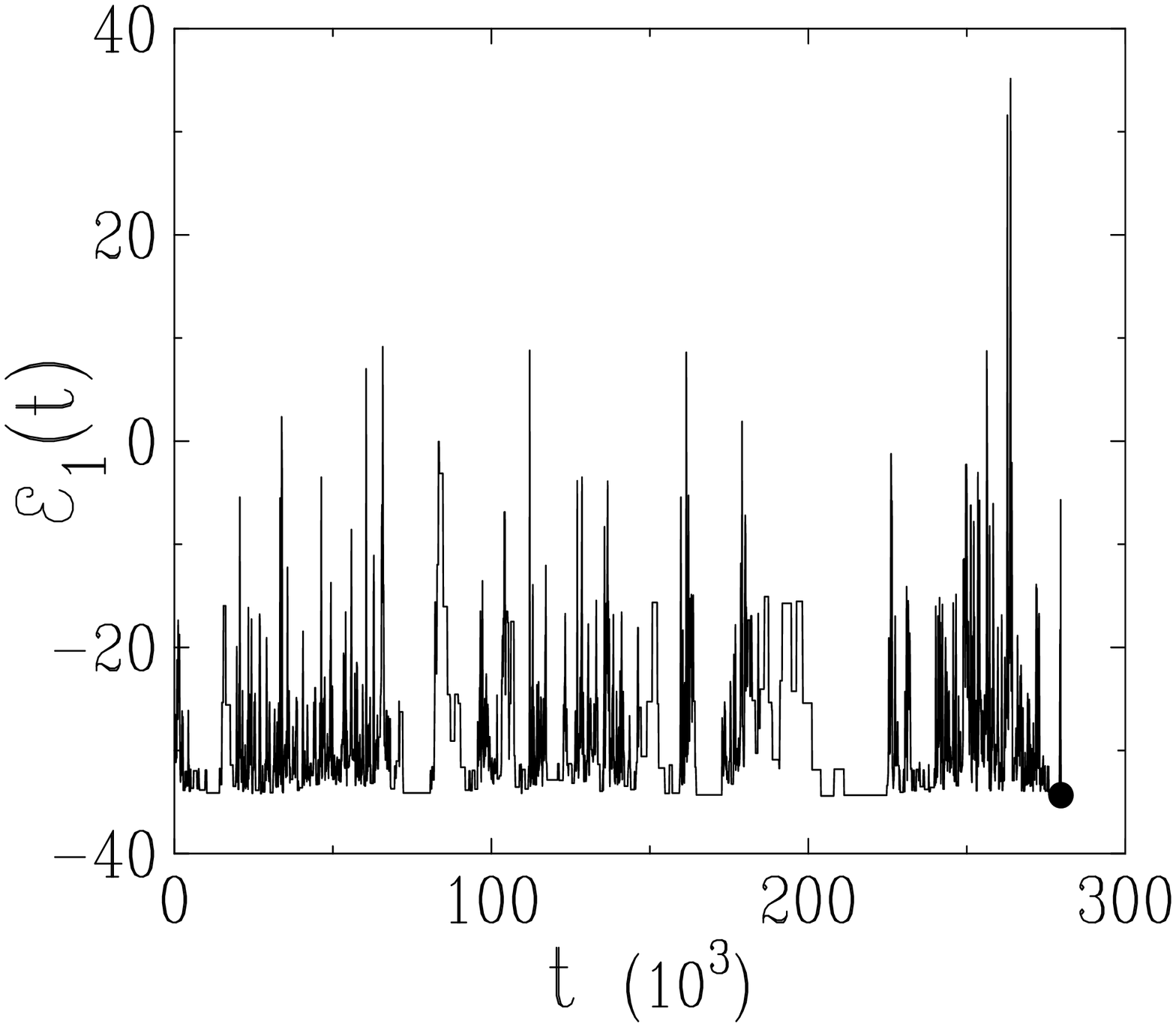}

\includegraphics[angle=90,height=4.5truecm]{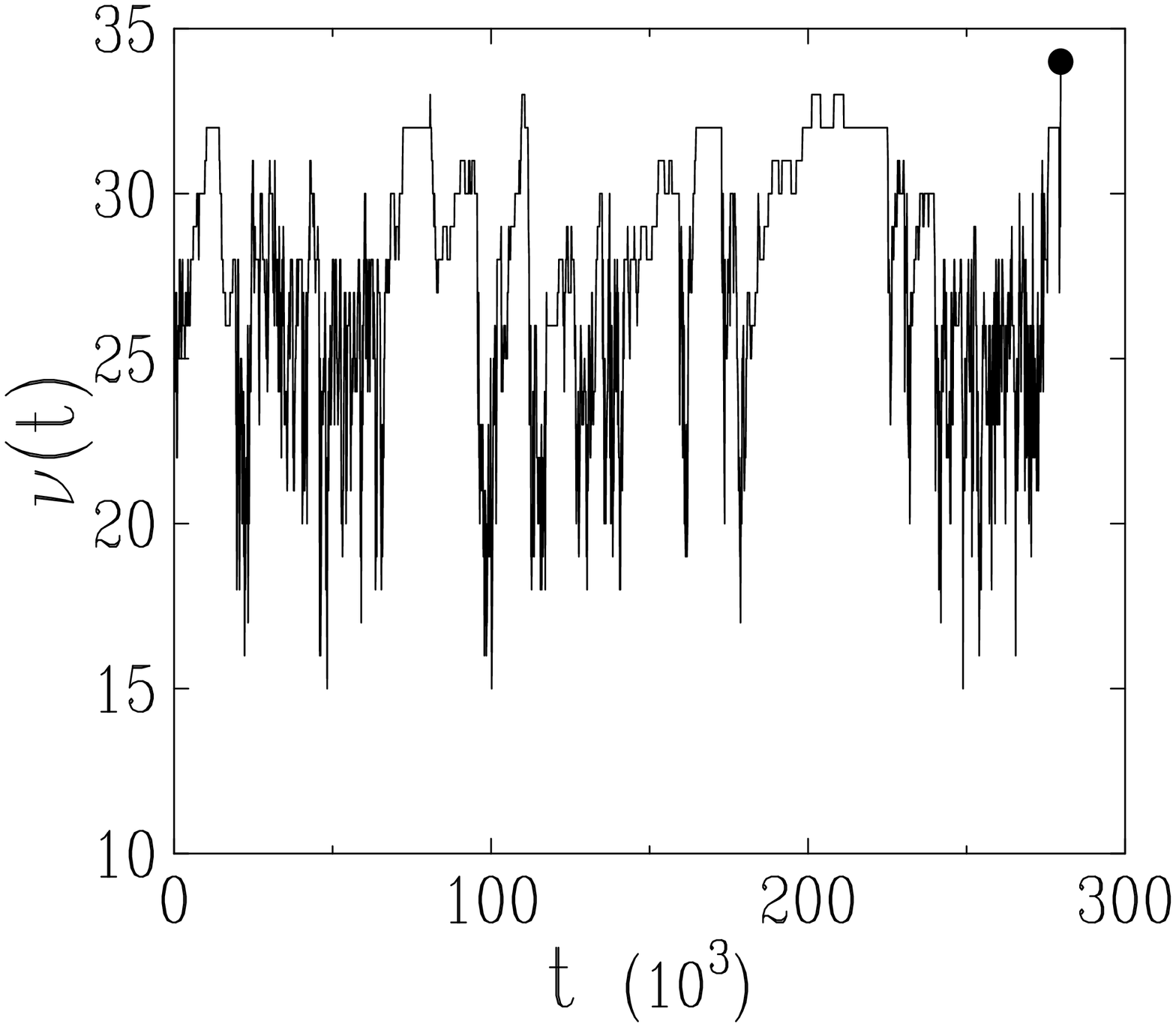}

\includegraphics[angle=90,height=4.5truecm]{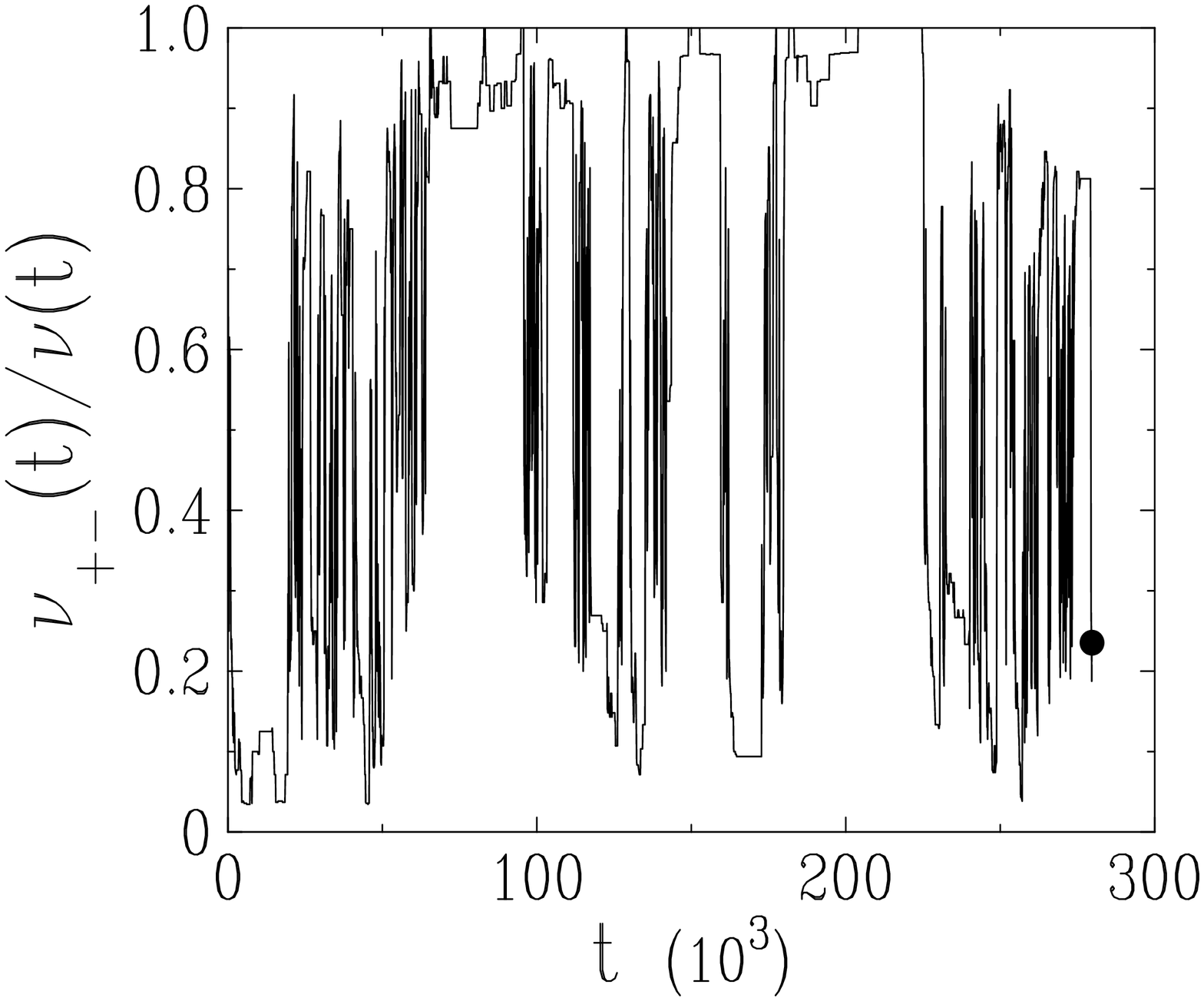}
\caption{\small
Top to bottom:
plots of the thickness $L(t)$ of the ordered upper layer,
the second component $\E_1(t)$ of the pseudo-energy,
the numbers $\nu(t)$ of dimers,
and the fraction $\nu_{+-}(t)/\nu(t)$ of $(+-)$ dimers,
for the history illustrating Phase IV in Figure~\ref{figabcd}.
Full symbols: values of the observables in the attractor,
i.e., right at the jamming time $T$.}
\label{figp}
\end{center}
\end{figure}

This phenomenon is illustrated in Figure~\ref{figp} from different viewpoints,
using the time dependence of four observables (for the same stochastic history
which was used to illustrate Phase IV in Figure~\ref{figabcd}).
The jamming time $T\approx279\,668$ for this history
is about 2.4 times larger than the mean jamming time for the parameters
$N=70$, $\xii=50$, $\xidy=7$.
The plotted observables are:

\noindent $\bullet$
the thickness $L(t)$ (see (\ref{ldef})) of the ordered upper layer,

\noindent $\bullet$
the second component $\E_1(t)$ (see (\ref{ecomp})) of the pseudo-energy,

\noindent $\bullet$
the total number $\nu(t)$ of dimers,

\noindent $\bullet$
the fraction $\nu_{+-}(t)/\nu(t)$ of $(+-)$ dimers.

The last two quantities involve the following definitions
of the numbers $\nu_{+-}$ of $(+-)$ dimers and $\nu_{-+}$ of $(-+)$ dimers:
\beq
\matrix{
\ds{\nu_{+-}=\quarter\sum_{k=2}^{N/2}(1+\s_{2k-1})(1-\s_{2k})},\hfill\cr
\ds{\nu_{-+}=\quarter\sum_{k=2}^{N/2}(1-\s_{2k-1})(1+\s_{2k})},\hfill\cr}
\label{nudefs}
\eeq
and of the total number $\nu=\nu_{+-}+\nu_{-+}$ of dimers
in a given configuration:
\beq
\nu=\nu_{+-}+\nu_{-+}=\half\sum_{k=2}^{N/2}(1-\s_{2k-1}\s_{2k}).
\eeq
The above formulae exclude the uppermost dimer,
which is fixed by the boundary condition (\ref{init}).

The four tracks shown in Figure~\ref{figp} all show
{\it strongly correlated, intermittent and non-stationary} fluctuations
at all time scales, ranging from the instantaneous to scales
of order of the jamming time $\mean{T}$ itself.
These features are commonly observed in glassy systems.
The existence of a glassy phase exhibiting this phenomenology
in a one-dimensional model represents
one of the most interesting outcomes of this work.

If we examine the dynamical history depicted in Figure~\ref{figp}, we will
notice that it can be described as an alternation between two
different kinds of periods:

\noindent$\bullet$ {\it Periods of quietude.}
Four such periods are visible in the figure.
They are characterised by quasi-stationary states with a high degree of order.
The thickness $L(t)$ and the number $\nu(t)$ of dimers fluctuate around
their maximal ground-state values of 70 and 34 respectively; the
pseudo-energy $\E_1(t)$ is correspondingly minimised,
and the fraction of $(+-)$ dimers is either close to zero or close to unity,
indicating a highly polarised column
which is close to one of the crystalline attractors $U_\pm$.
In some sense, it is as if the system has almost made its mind
up to choose one of the two global attractors, and is dawdling in its vicinity
with nearly no major fluctuations, during each of these periods of quietude.
However, these long excursions do not in any sense
anticipate the fate of the column.
In the given example,
the attractor finally chosen (full symbol) is close to $U_-$,
although the system spends most of its time in
the vicinity of the attractor $U_+$
with the fraction of $(+-)$ dimers typically close to unity during
the periods of quietude.

\noindent$\bullet$ {\it Itinerant periods.}
During these periods of confused wandering
between two consecutive periods of quietude,
all the indicators fluctuate wildly with no particular aim in sight.
All the observables monitored in Figure~\ref{figp}
are characterised by low order,
with the pseudo-energy even going positive on occasion.

We end up with a speculation based on a pictorial analogy.
The tracks in Figure~\ref{figp} are reminiscent of those obtained in
avalanche dynamics \cite{avalanches},
where periods of small random events give rise to large
system-size avalanches, which are known to be due to stress buildup and
release on the surface.
It is interesting, using this analogy,
to speculate whether the itinerant periods in Figure~\ref{figp} build
up unsustainable geometric disorder all along the column, which can only be
relieved by a systemic choice of a nearly ordered configuration (that is
close to an attractor), in which the column then lives, until disorder strikes
again in the form of the next itinerant period.

\subsubsection*{Mean jamming time}

We now focus on a quantitative analysis of several aspects of the glassy phase,
beginning with the mean jamming time $\mean{T}$.
Recall that in the activated phase, i.e., for $\xii>\xiic$ and $\xidy$ large
enough, the jamming time grows exponentially fast with $N$ (see (\ref{fact})).
We now examine the effect of decreasing $\xidy$, to cross over into the glassy
phase.

The main effect of this is that an increasingly broad
spectrum of local frequency scales $\o_n$ kicks in, to slow down
the stochastic behaviour of the activated phase.
Interestingly, a toy model
of an assembly of independent two-level systems is able to provide
a clue to this crossover.
Its details are presented in Appendix C, but the crucial feature is that
it has two regimes -- one where entropy dominates (`entropic'),
and the other which is dominated by the slowest of the local frequencies
(`slow').

The mean jamming time is in fact what would be most naively expected
from the above competition, that is, it is the greater of the two
times that would be generated:
\beq
\mean{T}\sim\max\,\Bigl(\exp(a(\xii)N),\;\exp(N/\xidy)\Bigr),
\label{sharp}
\eeq
where $\exp(a(\xii)N)$ is the jamming time of (\ref{fact})
in the $\xidy\to\infty$ limit,
while $1/\o_N=\exp(N/\xidy)$ is the slowest microscopic time scale of the
problem.
More specifically, this implies the following:

\noindent$\bullet$
In the {\it activated} phase, when $\xidy>1/a(\xii)$,
the result (\ref{fact}) for the mean jamming time in the $\xidy\to\infty$ limit
is essentially unchanged -- this corresponds
to the {\it entropic} phase of the toy model of Appendix C.

\noindent$\bullet$
In the {\it glassy} phase, i.e., for $\xidy<1/a(\xii)$,
the jamming time grows proportionally to $1/\o_N=\exp(N/\xidy)$ -- this
corresponds to the {\it slow} phase of the toy model of Appendix C.

From the above, one naturally expects there to be
a sharp transition in the $\xii-\xidy$ plane, along the line defined by
\beq
\xidy(\xii)=\frac{1}{a(\xii)},
\label{tpdef}
\eeq
shown qualitatively in Figure~\ref{figdiag}.
For the strongest correlations, as $\xii\to\infty$, the transition point
$\xidy(\xii)$ takes its minimum value
$\xidy(\infty)=1/a_\infty=2/(\ln 2)\approx2.8854$ (see (\ref{ainf})).
On the other hand, at the boundary of the ballistic/logarithmic
and activated regimes ($\xii\to\xiic^+$), (\ref{crita})
predicts a divergence of the transition point of the form
\beq
\xidy(\xii)\approx115/(\xii-\xiic).
\eeq

Despite huge finite-size effects,
our simulation data (shown as full symbols in Figure~\ref{figtim})
manifest the crossover described above.
The plateau in the left part of the data (full horizontal line)
yields the effective value $a_{\rm eff}\approx0.155$ for $\xii=\infty$ and
$N=50$.
This effective value is very far from the theoretical
asymptotic value $a_\infty$ (see (\ref{ainf})),
underlining the importance of finite-size effects.
However, and reassuringly for our analysis, the crossover
does indeed take place as predicted by (\ref{tpdef}),
at a value of $1/\xidy\approx a_{\rm eff}$ (full vertical line).

\subsubsection*{Statistics of attractors}

We now turn to the statistics of attractors in the glassy phase.
The question of what they are is easily addressed.
Recall that the ground states of the model for $\eps=1$ are the $2^\nu$
configurations made up of $\nu=N/2-1$ dimers, which satisfy
the boundary condition (\ref{init}).
By construction, these are the possible attractors of zero-temperature dynamics.

The next question, which relates to their {\it dynamical attainability},
is less easy to answer.
A precise formulation of this question is: What is the probability
$Q(\C)$ that the application of zero-temperature dynamics leaves the column
in a given attractor $\C$,
starting from a uniformly chosen random initial configuration?
Or, more physically: {\it how and where does a constrained system,
starting from random initial conditions, attain jamming?}
This question has held centre stage in theoretical \cite{compcoop,gl} and
experimental \cite{sidnature}
explorations of granular media and many other complex systems,
ever since Edwards postulated that
the entropic landscape of granular systems was flat \cite{edwards}.
Edwards' {\it flatness} hypothesis (in the strong sense)
implies that the attractors are sampled uniformly by the dynamics,
so that $Q(\C)$ is independent of the attractor $\C$,
and therefore equal to the reciprocal of the total number of attractors.

Our reason for introducing these issues at such a late stage
in this paper is that the statistics of attractors
are likely to be non-trivial only in the glassy phase.
All the other phases indeed manifest sufficiently stochastic behaviour that
one would expect the entropic landscape to be at least approximately flat.

A central quantity in this framework is therefore the dynamical entropy
\beq
S=-\sum_\C Q(\C)\ln Q(\C).
\eeq
In the case where the attractors are sampled uniformly,
according to Edwards' hypothesis,
the dynamical entropy assumes its maximal value:
\beq
S_\max=\nu\ln 2,
\label{sedw}
\eeq
where $\nu=N/2-1$.

Measuring entropies directly via numerical simulations is known
to be a very difficult task.
Instead, we resort to an inspired guess.
Since it seems likely that the crystalline attractors $U_\pm$
introduced in (\ref{unidef}) will play a special role in the dynamics,
we use them implicitly to define quantities
of interest on the attractors reached by the dynamics:

\noindent$\bullet$
A {\it global} indicator is provided
by the probability distribution $p(\nu_{+-})$ of the number of $(+-)$ dimers.
By using the dimer variables $\eta_k$ introduced in (\ref{gs1}),
the definitions (\ref{nudefs}) can be simplified as:
\beq
\nu_{+-}=\half\sum_{k=2}^{N/2}(1+\eta_k),\quad
\nu_{-+}=\half\sum_{k=2}^{N/2}(1-\eta_k),
\eeq
so that $\nu_{+-}+\nu_{-+}=\nu=N/2-1$.

If Edwards' hypothesis holds,
i.e., if the $2^\nu$ attractors are all equally likely to occur
as attractors, the distribution of $\nu_{+-}$ is binomial:
\beq
p(\nu_{+-})=\frac{1}{2^\nu}\bin{\nu}{\nu_{+-}}
=\frac{\nu!}{2^\nu(\nu_{+-})!(\nu_{-+})!}.
\label{binlaw}
\eeq
In such a binomial distribution, the extremal values $\nu_{+-}=0$ and
$\nu_{+-}=\nu$,
corresponding to the crystalline attractors $U_\pm$
(where all the dimers are of the same kind) are the least probable.
On the other hand, if the actual distributions obtained deviate from
the binomial law (\ref{binlaw}), this would indicate strongly
that all the attractors are {\it not} equally likely,
the entropic landscape is {\it not} flat,
and thus of course that Edwards' hypothesis does not hold.

\noindent$\bullet$
A {\it local} indicator of attractor structure is the correlation function
\beq
\chi_k=\frac{\mean{(\nu_{+-}-\nu_{-+})\eta_k}-1}{\nu-1}.
\eeq
This correlation measures the trend for the $k$-th dimer
to be aligned with every other dimer.
It vanishes identically if Edwards' flatness hypothesis holds.
In the extreme opposite situation where only the crystalline
attractors $U_\pm$ are reached by the dynamics,
the above correlation takes its maximal value $\chi_k=1$ for all $k$.

\begin{figure}[!t]
\begin{center}
\includegraphics[angle=90,height=5truecm]{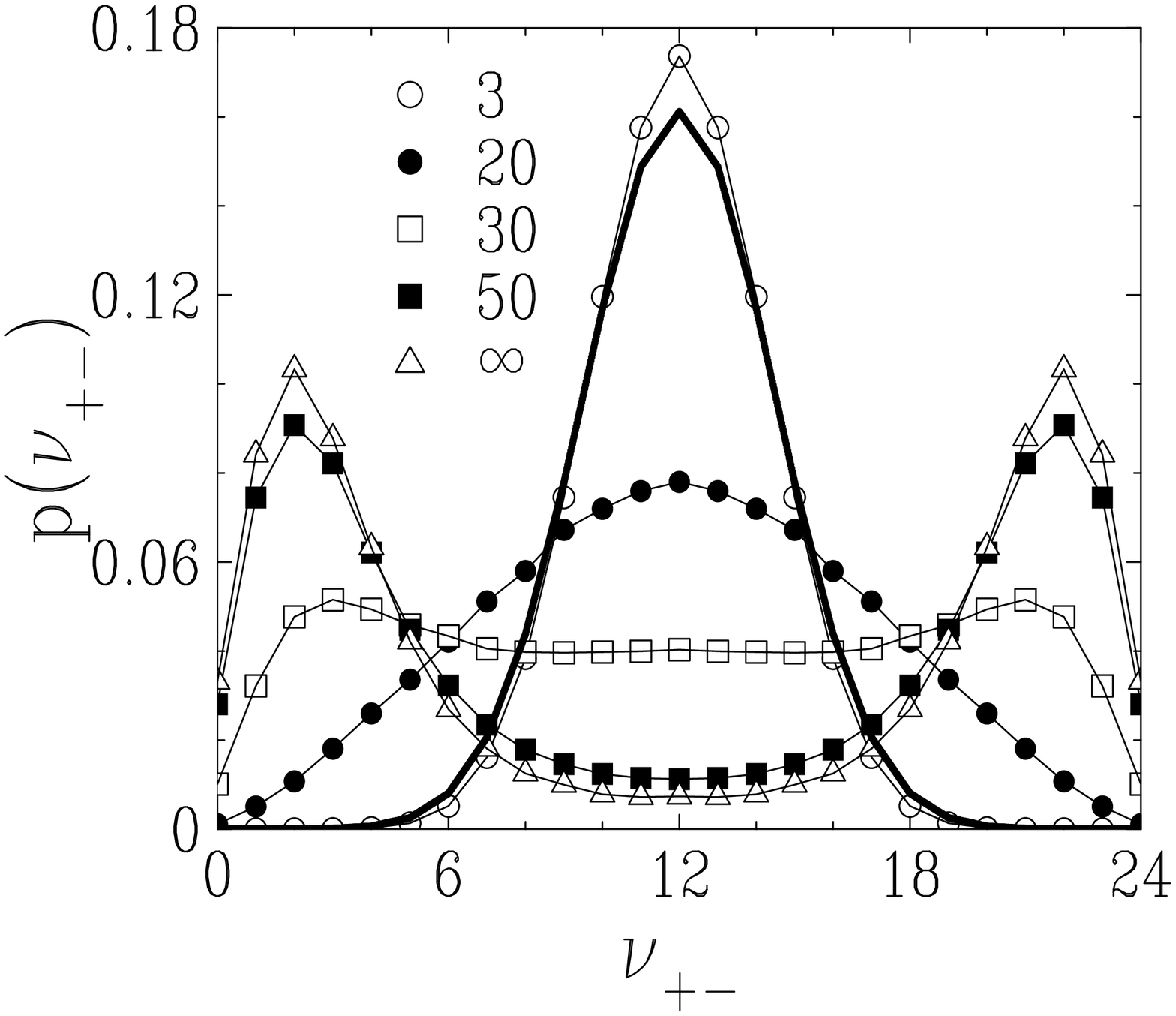}

\includegraphics[angle=90,height=5truecm]{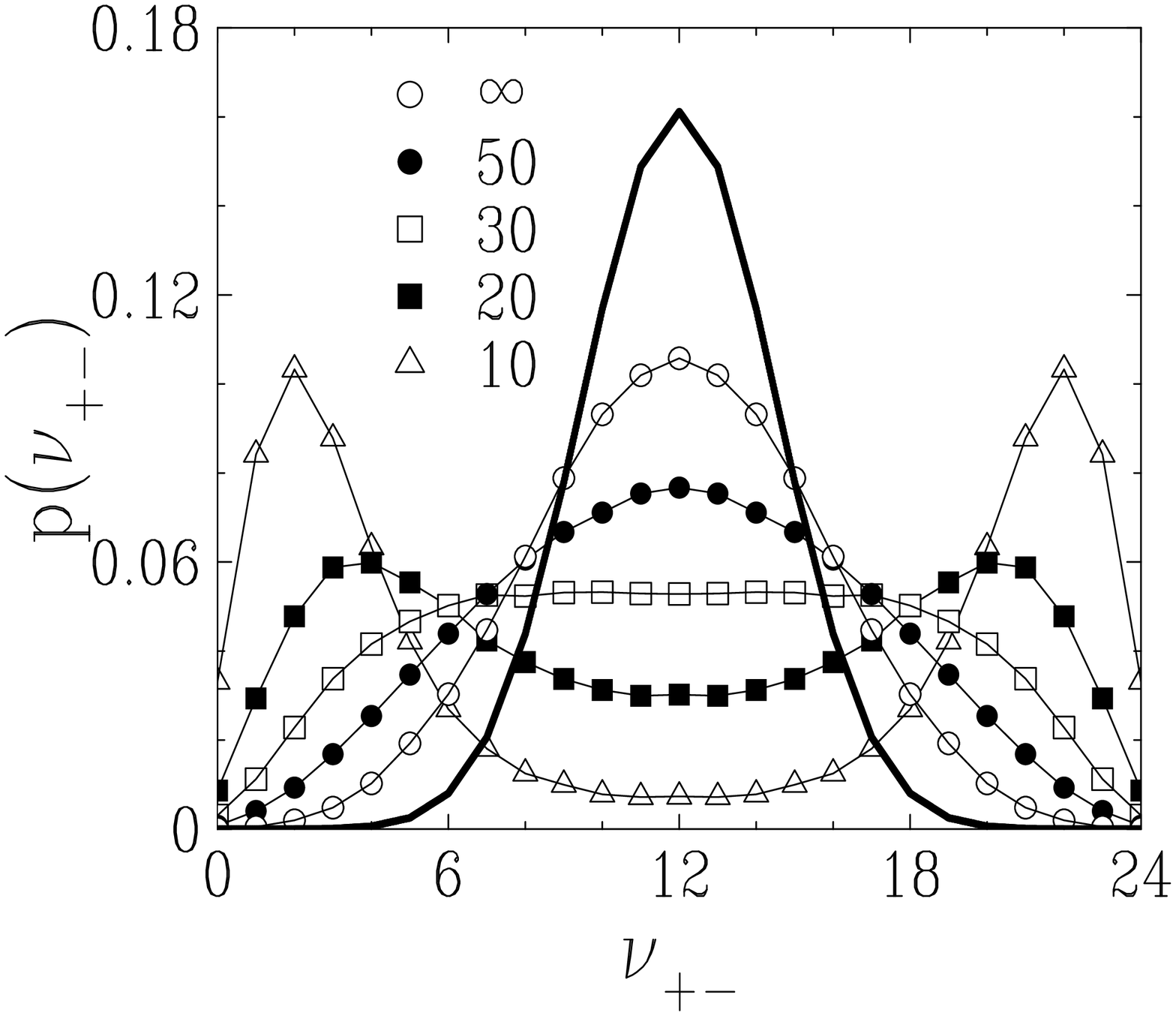}
\caption{\small
Histogram plots of the probability distribution $p(\nu_{+-})$
for $N=50$ (hence $\nu=24$).
Top: $\xidy=10$ and variable $\xii$.
Bottom: $\xii=\infty$ and variable $\xidy$.
The binomial distribution (\ref{binlaw}) is shown as thick full lines.}
\label{fig2}
\end{center}
\end{figure}

\begin{figure}[!t]
\begin{center}
\includegraphics[angle=90,height=5truecm]{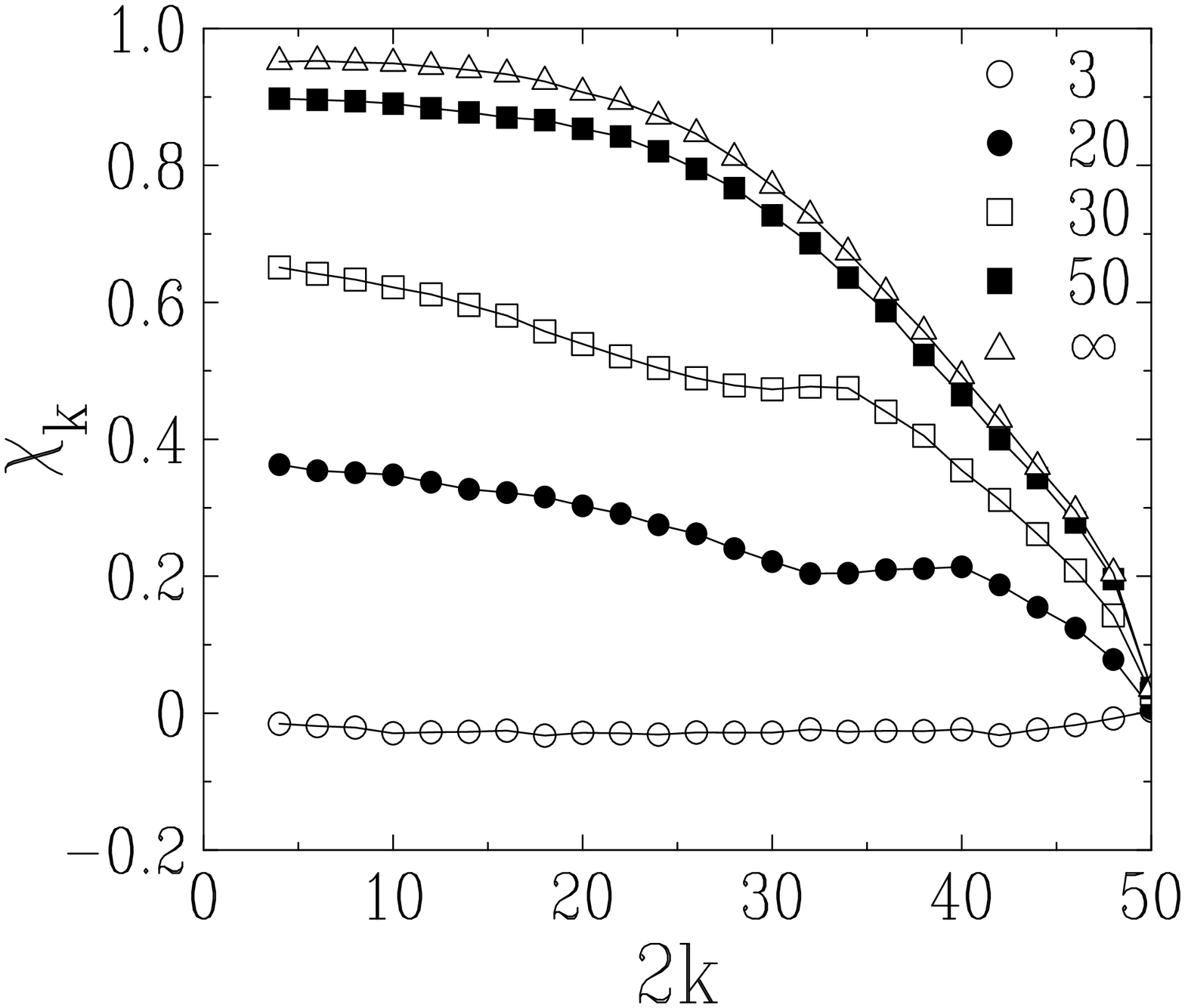}

\includegraphics[angle=90,height=5truecm]{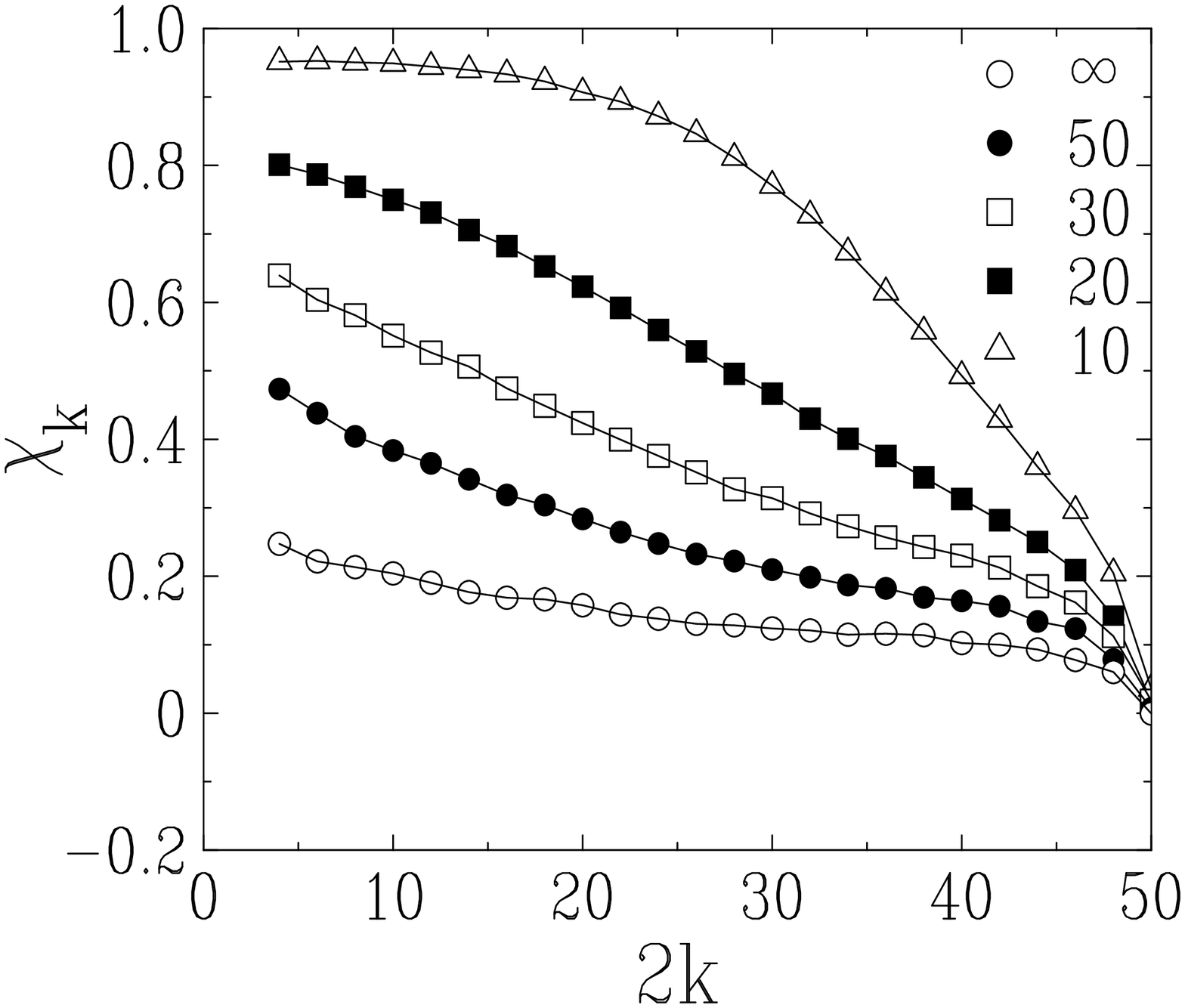}
\caption{\small
Plot of the dimer correlation function $\chi_k$ against depth $n=2k$.
Parameters are as in Figure~\ref{fig2}.}
\label{fig3}
\end{center}
\end{figure}

Our numerical results for the probability distribution $p(\nu_{+-})$
and the dimer correlation function
$\chi_k$ are shown in Figures~\ref{fig2} and \ref{fig3}.
All the data were taken for a system of size $N=50$,
which has $\nu=24$ dimers that are free to reorient.
Figure~\ref{fig2} shows the variation
of the form of $p(\nu_{+-})$ with, first, fixed $\xidy=10$ and variable
$\xii$, and next, fixed $\xii=\infty$ and variable $\xidy$.
Figure~\ref{fig3}
shows the variation of $\chi_k$ along the same diagnostic lines.

In Figure~\ref{fig2}, the binomial distribution (corresponding to Edwards'
flatness hypothesis) is shown by a thick full line, with which the
data for the lowest value of $\xii$ are
almost completely aligned.
The statistics of attractors is thus very close to being uniform
in the ballistic and logarithmic phases, which correspond to the top layers
of a column.
As $\xii$ increases, there is a gradual crossover
to a non-trivial two-peaked distribution;
the same trend is visible in the lower panel,
when $\xidy$ decreases for infinite $\xii$.
At the beginning of the crossover,
with its small deviations from uniform sampling, one recognises
the activated phase, which corresponds to the middle of a column.
By the time that the two-peaked distribution is obtained
in both parts of Figure~\ref{fig2}, the parameters -- $\xii$ large, and
$\xidy$ small -- correspond clearly to the glassy phase.
Here, attractors in the neighbourhoods
of the crystalline states $U_\pm$ are eventually
favoured, after long periods of systemic wandering.

The above observations are reinforced by Figure~\ref{fig3}.
In the upper panel,
the correlation function $\chi_k$ is essentially zero for low $\xii$,
increasing progressively as $\xii$ is increased.
In the lower panel, i.e., for infinite $\xii$,
the correlation function is never quite zero even for high $\xidy$, and
only manifests a stronger depth-dependence as $\xidy$ decreases.
These results reinforce those found in an independent model which
uses random graphs to model grains near jamming, where entropic deviations
from Edwards' flatness occur in certain regions of parameter space
\cite{johannes}.

To recapitulate, the salient feature that emerges is that the
system prefers increasingly to live in the neighbourhood of its two global
minima, the attractors $U_\pm$, as one goes deep into the glassy phase.
As a consequence, the dynamical entropy decreases from a value
close to its maximal value (\ref{sedw})
at the boundary between Phases II and IV,
to zero in the deepest part of the glassy phase ($\xii\gg N$, $\xidy\ll N$).
Furthermore, Edwards' flatness is well obeyed overall in three out of
four phases in this model; it is massively violated only
deep in the glassy phase of this model, where the configurational
landscape is completely rough.

Our success in constructing this glassy phase via such a minimal model
relied on the inclusion of two crucial ingredients:

\noindent$\bullet$
Long-range interactions ($\xii>N$), in order for the dimers to jam
cooperatively, rather than independently of each other;

\noindent$\bullet$
A broad spectrum of local frequencies ($\xidy$ small) to slow down
the relaxation, and thus prevent a purely activated mechanism driven by entropy.

It is noteworthy that both these minimal ingredients rely on {\it collective}
effects -- one to do with interactions in {\it space}, the other
to do with degrees of freedom in {\it time}.
This model-specific conclusion
at once agrees with, and reinforces, general
notions \cite{miguel} of cooperativity in glassy dynamics.

\section{Discussion}
\setcounter{equation}{0}
\def\theequation{5.\arabic{equation}}

The motivation for this model came from a mental image of grains
in a box under shaking: what
could be at once so simple, or -- as we came to see in time -- so complex? Our
first, simplest, model \cite{I,II} involved only the effects of gravity
on non-interacting grains -- deeper grains carried the weight of grains above
them, so were less free to move.
This was modelled by using a single dynamical length $\xidy$,
representing the thickness of the dynamical boundary layer:
grains at a depth $n$ much less than $\xidy$
(which can be examined by setting $\xidy\to\infty$) are free to move, whereas
those where $n\sim\xidy$ have lower frequencies of motion, as normal in
non-Newtonian fluids.
Three phases were found; in the `fluidised' phase,
grains flew as well as moved along the surface, with a relatively quick
propagation of order down the sandbox.
Grain disorder was essentially frozen in, in the `glassy' phase, with
a very slow propagation of order from the free surface.
The `intermediate' phase
was in some ways the most interesting, with a true competition between
fast and slow dynamics.

In hindsight, it is astonishing that these diverse behaviours
-- especially the shape-dependent `ageing' effects of the glassy regime --
were manifested in a totally non-interacting model.
A more realistic, interacting model
of the glassy regime was presented in \cite{III,IV}.
Since close-packed grains can typically not diffuse spatially, it was
sufficient to model a column, rather than a box, of grains.
In the model of \cite{III,IV}, grain motions were constrained not just
by the masses, but also by the orientations of grains above them,
thus generating directional long-range interactions.
The effect of compaction around jamming was modelled by a single
local field $h_n$, representing the excess void space \cite{br}
for grain $n$, which could be minimised
by a suitable choice of grain orientation.
Additionally, in this model,
grains were allowed to have arbitrary shapes, so that the
the disordered orientation of a grain could occupy any volume $\eps$,
and, correspondingly generate any void space.
The propagation of order in this model proceeded from the free surface to the
base, and was `causal in space' --
in that while upper grains constraint lower ones, the converse was not true.

It took us some time and several explorations to realise that while the
model of \cite{III,IV} had at least the flavour of the interactions needed
to model a jammed glassy phase -- e.g. the constraining effect of long-range
grain correlations -- its
lack of slow dynamics (except those arising from the trivial effect
of grain masses) was a direct result of its spatial causality.
Essentially, provided a grain was not blocked down by the weight of other
grains, it was free to orient itself subject only to the orientations
of grains {\it above} itself -- that is, we were modelling the behaviour of the
top layers of a jammed column of grains, which never felt the undertow of the
base.
It was small surprise, therefore, that the ordering dynamics
for $\xidy\to\infty$ were ballistic.

To model a column of grains with spatial inhomogeneities -- that
is, a column with a top, a middle, and a bottom --
we discovered that orientational constraints needed to be inserted in a
{\it non-directed} way.
This enabled us to model frustration -- in this context, the need of a given
grain to balance
the effects of two competing local fields $h_n$ and $j_n$ -- which led
in its turn, to slow dynamics.
Still keeping
the orientational constraints of previous models \cite{I,II,III,IV} via
the field $h_n$, as well as the effect of `gravitational slowing down' via
$\xidy$,
we therefore introduced here the notion that grains were also constrained
by grains {\it below} them, via the field
$j_n$ which propagated over a correlation length $\xii$.

From this very heuristic and pictorial modelling
has emerged a model column of grains that manifests all the complexity
of earlier models \cite{III,IV}, and adds
some more via the introduction of the activated and glassy phases.
Our main success is of course in the realisation of a glassy phase
for which a minimal combination of two physical
ingredients -- {\it strong, bi-directional, orientational correlations and a
broad spectrum of local frequencies arising from a natural depth-dependence} --
appears to be necessary.
The richness of this phase is worthy of further
exploration, especially to do with issues concerning higher-order correlations
and ageing.

Finally, we mention that our model provides
some rather interesting and general insights into the nature
of optimisation -- the granular column modelled
here reaches its ground states in
strikingly different ways, in the four dynamical phases mentioned above.
It is tempting to think of these phases as representing different spatial
parts -- `top', `middle' and `bottom' -- of a column, and to connect their
different routes to compaction with the issue of
inhomogeneities in real granular media \cite{0}.
While this picture is an
appealing one, one should remember that the four phases of this model
were obtained by varying $\xii$ and $\xidy$; the translation of our results
to apply to a real column
would involve the natural apparition of such variations as a function of depth.

\appendix
\section*{Appendix A. Exact results for $N=4$ and $N=6$}
\setcounter{equation}{0}
\def\theequation{A.\arabic{equation}}

In this Appendix we derive analytical results
on the zero-temperature dynamics of the model investigated
in the body of the paper, for small systems.
We concentrate onto the statistics of attractors,
i.e., the probability that the system is absorbed in each attractor,
starting from a random initial configuration.
Exact results will be successively obtained for systems of sizes $N=4$
and $N=6$, for arbitrary $\xii$ and $\xidy$.

\subsubsection*{The case $N=4$}

The column made of 4 grains with the boundary condition (\ref{init})
has three free orientations
: $\s_2\pm1$, $\s_3=\pm1$, $\s_4=\pm1$,
and therefore eight configurations, in lexicographical order:
\beq
\matrix{
\C_1=++++,\quad\C_2={+++-},\quad\C_3=++-+,\hfill\cr
\C_4=++--,\quad\C_5={+-++},\quad\C_6=+-+-,\hfill\cr
\C_7=+--+,\quad\C_8={+---}.\hfill}
\eeq

The two attractors are the dimerised configurations $\C_6$ and $\C_7$,
respectively corresponding to $\eta_2=+1$ and $\eta_2=-1$.
Our main goal is to determine the probabilities $Q^{(6)}$ and $Q^{(7)}$
that the system is absorbed in each of these attractors,
starting from a random initial configuration.
The relevant information is encoded in the non-uniformity parameter
\beq
\Delta=\mean{\eta_2}=Q^{(6)}-Q^{(7)}.
\eeq

The zero-temperature dynamics consists of a certain number of moves
between configurations.
For instance, $\C_1$ may be updated into the following configurations
at the following rates:
\beq
\C_1\react{\longrightarrow}{\o_2}\C_5,\quad
\C_1\react{\longrightarrow}{\o_3}\C_3,\quad
\C_1\react{\longrightarrow}{\o_4}\C_2.
\eeq
As announced in Section 3.3, this dynamics is independent of $\xii$ for $N=4$.
It can be represented as an $8\times8$ Markov matrix $\M$,
such that the occupation probabilities $P_a(t)$ of the configurations $\C_a$
($a=1,\dots,8$) obey the {\it forward Kolmogorov equation} \cite{redner,feller}
\beq
\frac{\d P_a(t)}{\d t}=\sum_b\M_{ab}P_b(t).
\label{kfe}
\eeq

The absorption probabilities of the attractors
can be derived by means of the following approach.
Let $Q^{(c)}_a$ be the probability of being eventually absorbed
by configuration $\C_c$, starting from the initial configuration $\C_a$.
For a fixed configuration $\C_c$, the absorption probabilities $Q^{(c)}_a$ obey
the {\it backward Kolmogorov equation} \cite{redner,feller}
\beq
\sum_a Q^{(c)}_a\M_{ab}=0,
\label{kbe}
\eeq
complemented by the boundary condition $Q^{(c)}_c=1$.
For a random initial configuration, we have therefore
\beq
Q^{(c)}=\frac{1}{8}\sum_a Q^{(c)}_a.
\eeq

By solving (\ref{kbe}) successively for both attractors ($c=6$ and $c=7$),
we are left with the following explicit result, for arbitrary rates $\o_n$:
\beq
\Delta=\frac{\o_3\o_4(\o_3-\o_4)}{2(\o_2+\o_3)(\o_2+\o_4)(\o_3+\o_4)}.
\eeq

In the present model, where $\o_n=\exdy^n$ (see (\ref{odef})),
this result simplifies to
\beq
\Delta=\frac{(1-\exdy)\exdy^3}{2(1+\exdy)^2(1+\exdy^2)}.
\label{delta4}
\eeq
The non-uniformity parameter $\Delta$ is a small positive quantity,
meaning that the crystalline attractor $\C_6=U_+$
is always slightly favoured by the dynamics.
It coincides with $\Delta_1$ plotted in the upper panel of Figure~\ref{fig6}.
It vanishes in both limits $\exdy\to0$ (i.e., $\xidy\to0$)
and $\exdy\to1$ (i.e., $\xidy\to\infty$).
Its maximum $\Delta\approx0.012465$ is reached for $\exdy\approx0.6253$.

\subsubsection*{The case $N=6$}

The case of a column made of $N=6$ grains can still be dealt with
by analytical means, although the final expressions are much lengthier.
The system has 32 configurations.
Its four attractors are the following configurations
(relabelled as $a=1,\dots,4$ for convenience):
\beq
\matrix{
\C_1=+-+-+-,\quad\C_2=+-+--+,\hfill\cr
\C_3=+--++-,\quad\C_4=+--+-+.\hfill}
\eeq
The relevant information is encoded in the three non-uniformity parameters
\beq
\matrix{
\Delta_1=\mean{\eta_2}
=Q^{(1)}+Q^{(2)}-Q^{(3)}-Q^{(4)},\hfill\cr
\Delta_2=\mean{\eta_3}
=Q^{(1)}-Q^{(2)}+Q^{(3)}-Q^{(4)},\hfill\cr
\Delta_3=\mean{\eta_2\eta_3}
=Q^{(1)}-Q^{(2)}-Q^{(3)}+Q^{(4)}.\hfill}
\eeq

As announced in Section 3.3, $N=6$ is the smallest system size
such that the dynamics depends in a non-trivial way on the parameter $\xii$.
The Markov matrix $\M$ assumes two different expressions
for $0<\exi<\phi$ and $\phi<\exi<1$, where the inverse golden mean
$\phi$ is given in (\ref{phidef}).
For each of these two phases,
the $32\times32$ Markov matrix $\M$ has been generated,
and the backward equations (\ref{kbe}) corresponding to each of the four
attractors have been solved analytically.\footnote{with help
of the software MACSYMA.}
We thus obtain the following expressions for the non-uniformity parameters,
in the range $0<\exi<\phi$:
\beq
\matrix{
\Delta_1=\frac{(1-\exdy)\exdy^3}{2(1+\exdy)^2(1+\exdy^2)},\hfill\cr
\Delta_2=(\exdy^4-1)(\exdy^3+\exdy^2+1)\hfill\cr
{\hskip 5truemm}\times(\exdy^4+\exdy^3+\exdy^2+1)\exdy^6
\frac{P_{24}(\exdy)}{D(\exdy)},\hfill\cr
\Delta_3=-(\exdy-1)^2(\exdy^2+1)(\exdy^3+\exdy^2+1)\hfill\cr
{\hskip 5truemm}\times(\exdy^4+\exdy^3+\exdy^2+1)\exdy^5
\frac{P_{25}(\exdy)}{D(\exdy)},\hfill}
\label{for6inf}
\eeq
and in the range $\phi<\exi<1$:
\beq
\matrix{
\Delta_1=-\exdy^3\frac{P_{38}(\exdy)}{D(\exdy)},\hfill\cr
\Delta_2=(\exdy^4-1)(\exdy^3+\exdy^2+1)\hfill\cr
{\hskip 5truemm}\times(\exdy^4+\exdy^3+\exdy^2+1)\exdy^6
\frac{P_{24}(\exdy)}{D(\exdy)},\hfill\cr
\Delta_3=(1-\exdy^4)(\exdy^3+\exdy^2+1)\hfill\cr
{\hskip 5truemm}\times(\exdy^4+\exdy^3+\exdy^2+1)
\frac{P_{30}(\exdy)}{D(\exdy)},\hfill}
\label{for6sup}
\eeq
where we have introduced the following polynomials:
\beq
{\hskip -3.8truemm}\matrix{
{\hskip 2.3truemm}D(x)=8(x+1)^3(x^2+1)^2(x^2-x+1)(x^2+x+1)^2\hfill\cr
{\hskip 10.3truemm}\times(x^3+x+1)^2(x^3+x^2+1)^2(x^4+x^2+x+1)\hfill\cr
{\hskip 10.3truemm}\times(x^4+x+1)(x^4+x^3+1)(x^4+x^3+x^2+1),\hfill}
\eeq
\beq
{\hskip -5.3truemm}\matrix{
P_{24}(x)=4x^{24}+5x^{23}+10x^{22}+15x^{21}+9x^{20}-5x^{19}\hfill\cr
{\hskip 10.3truemm}-51x^{18}-104x^{17}-203x^{16}-308x^{15}-434x^{14}\hfill\cr
{\hskip 10.3truemm}-515x^{13}-605x^{12}-634x^{11}-609x^{10}-533x^{9}\hfill\cr
{\hskip 10.3truemm}-449x^{8}-330x^{7}-218x^{6}-130x^{5}-71x^{4}\hfill\cr
{\hskip 10.3truemm}-27x^{3}-3x^{2}+x+1,\hfill}
\eeq
\beq
\matrix{
P_{25}(x)=6x^{25}+15x^{24}+40x^{23}+95x^{22}+185x^{21}\hfill\cr
{\hskip
10.3truemm}+335x^{20}+547x^{19}+842x^{18}+1177x^{17}+1556x^{16}\hfill\cr
{\hskip 10.3truemm}+1900x^{15}+2193x^{14}+2349x^{13}+2356x^{12}\hfill\cr
{\hskip 10.3truemm}+2223x^{11}+1955x^{10}+1609x^{9}+1242x^{8}+888x^{7}\hfill\cr
{\hskip 10.3truemm}+590x^{6}+367x^{5}+207x^{4}+101x^{3}+43x^{2}\hfill\cr
{\hskip 10.3truemm}+17x+4,\hfill}
\eeq
\beq
{\hskip -3.5truemm}\matrix{
P_{30}(x)=6x^{30}+21x^{29}+54x^{28}+140x^{27}+292x^{26}\hfill\cr
{\hskip 10.3truemm}+546x^{25}+949x^{24}+1531x^{23}+2283x^{22}\hfill\cr
{\hskip 10.3truemm}+3207x^{21}+4277x^{20}+5377x^{19}+6457x^{18}\hfill\cr
{\hskip 10.3truemm}+7343x^{17}+8025x^{16}+8316x^{15}+8236x^{14}\hfill\cr
{\hskip 10.3truemm}+7793x^{13}+7011x^{12}+5978x^{11}+4852x^{10}\hfill\cr
{\hskip 10.3truemm}+3730x^{9}+2682x^{8}+1811x^{7}+1142x^{6}+662x^{5}\hfill\cr
{\hskip 10.3truemm}+343x^{4}+160x^{3}+64x^{2}+20x+4,\hfill}
\eeq
\beq
{\hskip -0.5truemm}\matrix{
P_{38}(x)=4x^{38}+28x^{37}+109x^{36}+341x^{35}+909x^{34}\hfill\cr
{\hskip 10.3truemm}+2114x^{33}+4416x^{32}+8424x^{31}+14836x^{30}\hfill\cr
{\hskip 10.3truemm}+24262x^{29}+37105x^{28}+53309x^{27}+72124x^{26}\hfill\cr
{\hskip 10.3truemm}+92105x^{25}+111177x^{24}+127001x^{23}\hfill\cr
{\hskip 10.3truemm}+137190x^{22}+140074x^{21}+135038x^{20}\hfill\cr
{\hskip 10.3truemm}+122612x^{19}+104406x^{18}+82925x^{17}+60963x^{16}\hfill\cr
{\hskip 10.3truemm}+40790x^{15}+24131x^{14}+11829x^{13}+3799x^{12}\hfill\cr
{\hskip 10.3truemm}-720x^{11}-2614x^{10}-2891x^{9}-2404x^{8}-1693x^{7}\hfill\cr
{\hskip 10.3truemm}-1031x^{6}-555x^{5}-269x^{4}-112x^{3}-39x^{2}\hfill\cr
{\hskip 10.3truemm}-11x-2.\hfill}
\eeq

\begin{figure}[!t]
\begin{center}
\includegraphics[angle=90,height=5truecm]{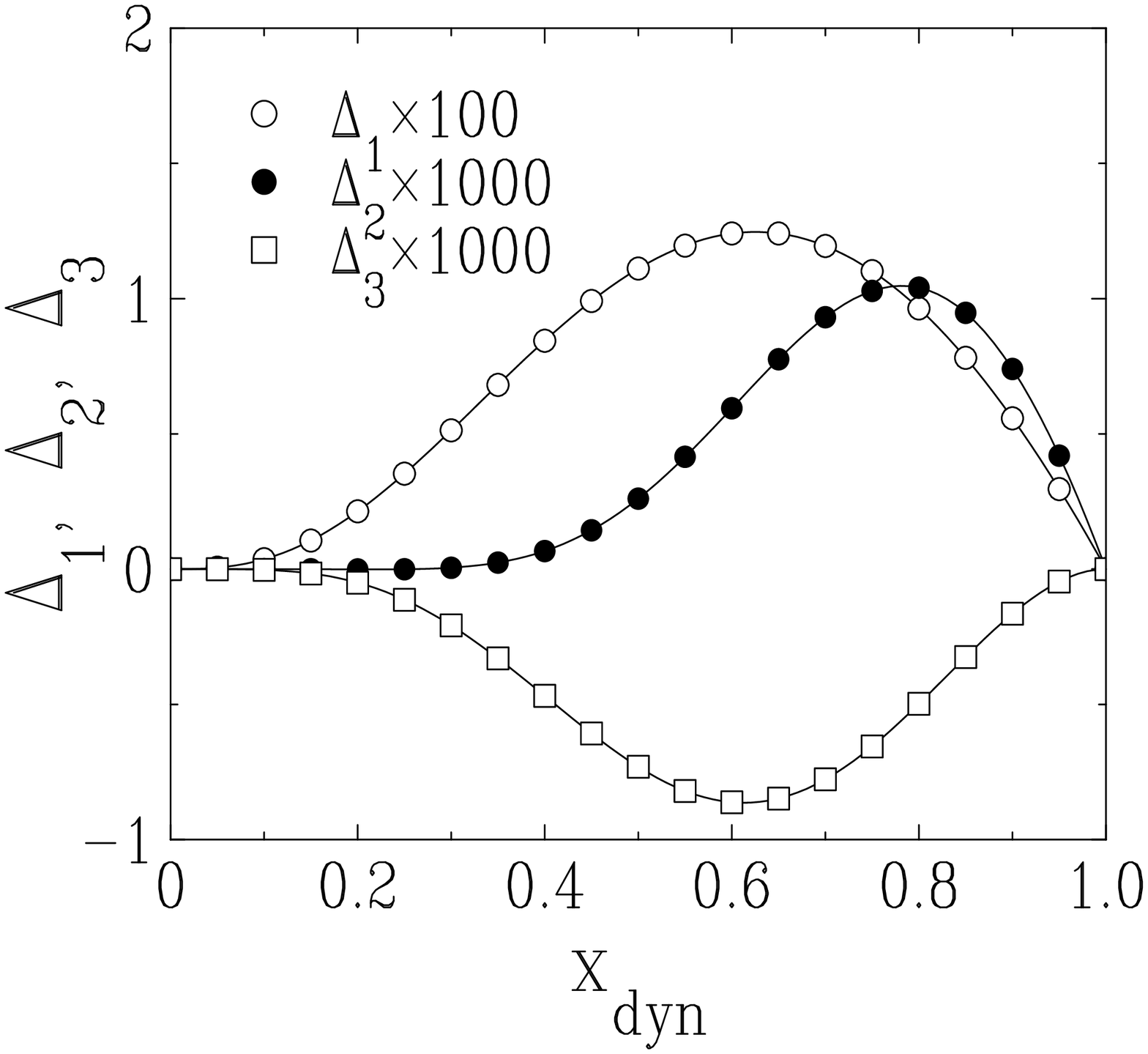}

\includegraphics[angle=90,height=5truecm]{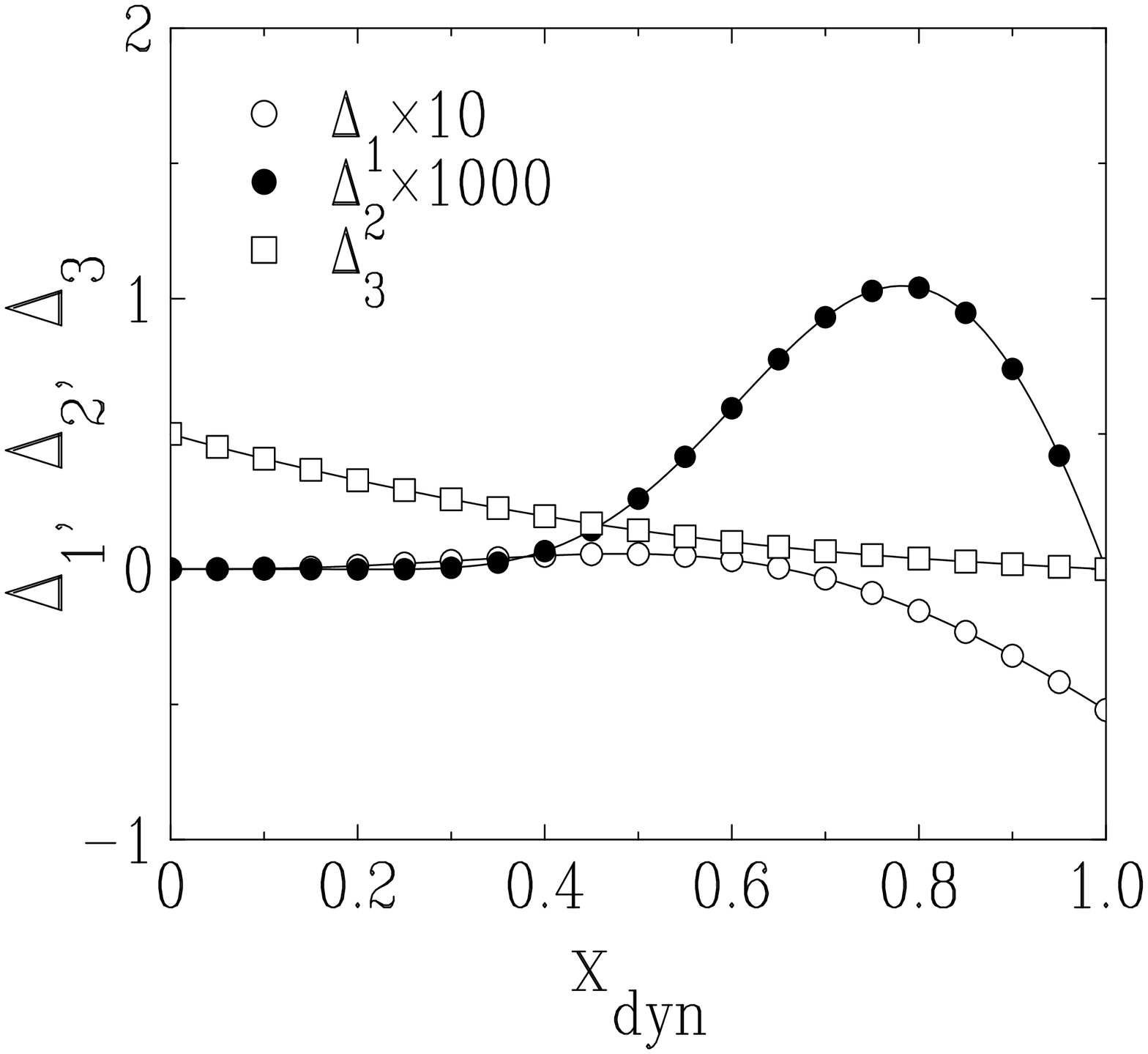}
\caption{\small
Plots of the non-uniformity parameters $\Delta_i$ ($i=1,2,3$)
characterising the attractor statistics of a column of $N=6$ grains.
Top: $0<\exi<\phi$ (see (\ref{for6inf})).
Bottom: $\phi<\exi<1$ (see (\ref{for6sup})).
Note the powers of 10 in the vertical scales.}
\label{fig6}
\end{center}
\end{figure}

The non-uniformity parameters $\Delta_i$ are plotted in Figure~\ref{fig6}
(note the powers of 10 in the vertical scales).
For $0<\exi<\phi$ (upper panel), the $\Delta_i$ are typically small.
They vanish in both limits $\exdy\to0$ and $\exdy\to1$.
$\Delta_1$ is always positive,
and it coincides with $\Delta$ of the case $N=4$ (see (\ref{delta4}));
$\Delta_2$ turns from negative to positive for $\exdy\approx0.2598$,
whereas $\Delta_3$ is always negative.
For $\phi<\exi<1$ (lower panel), the $\Delta_i$ are typically larger.
Only $\Delta_2$ is the same in both phases.
The $\Delta_i$ again vanish in both limits $\exdy\to0$ and $\exdy\to1$,
except for $\Delta_1=-5/96$ for $\exdy=1$ and $\Delta_3=1/2$ for $\exdy=0$.
Finally, $\Delta_1$ turns from positive to negative for $\exdy\approx0.6581$,
whereas $\Delta_3$ is always positive.

\section*{Appendix B. Biased Brownian motion on an interval}
\setcounter{equation}{0}
\def\theequation{B.\arabic{equation}}

In this appendix we consider biased Brownian motion on the interval $0<x<L$,
characterised by its drift velocity $V$ and its diffusion coefficient $D$.
The endpoint at $x=0$ is reflecting, whereas the endpoint at $x=L$ is absorbing.

This mixed type of boundary conditions is referred to
as the {\it transmission mode} \cite{redner}.
It ensures that, starting at a given position $x=\ell$,
the particle hits the absorbing endpoint with certainty
in some finite random time $T$, referred to as the absorption time.
Our main goal is to determine the distribution $\rho(T)$
of this absorption time, and especially its first few moments.
We shall mostly concentrate onto the $\ell\to0$ limit,
where the particle starts from the immediate vicinity of the reflecting
endpoint.

Let $P(x,t)$ be the probability density for the particle to be at point $x$
at time $t$, and $J(x,t)$ the associated probability current.
We have
\beq
\frac{\dpar P}{\dpar t}+\frac{\dpar J}{\dpar x}=0,\quad
J=VP-D\,\frac{\dpar P}{\dpar x},
\eeq
with the initial and boundary conditions
\beq
P(x,0)=\delta(x-\ell),\quad
J(0,t)=P(L,t)=0.
\eeq
Introducing the Laplace transform
\beq
\w P(x,s)=\int_0^\infty P(x,t)\,\e^{-st}\,\d t,
\label{lapdef}
\eeq
the above equations imply
\beq
s\w P+V\,\frac{\dpar\w P}{\dpar x}-D\,\frac{\dpar^2\w P}{\dpar x^2}
=\delta(x-\ell),
\label{pbulk}
\eeq
with boundary conditions
\beq
V\w P(0,s)-D\,\frac{\dpar\w P(0,s)}{\dpar x}=\w P(L,s)=0.
\label{pbound}
\eeq
Equation (\ref{pbulk}) implies the matching conditions
\beq
\matrix{
\w P(\ell^+,s)=\w P(\ell^-,s),\hfill\cr
\frac{\dpar\w P(\ell^+,s)}{\dpar x}-\frac{\dpar\w P(\ell^-,s)}{\dpar
x}=-\frac{1}{D}.}
\label{pmatch}
\eeq

Consider first the homogeneous equation obtained by setting
the right-hand side of (\ref{pbulk}) equal to zero.
Looking for a solution to this equation at fixed $s$ of the form
$\w P=\e^{rx}$,
we obtain the quadratic equation $s+Vr-Dr^2=0$, whose two roots read
\beq
r_1=\frac{V-W}{2D},\quad r_2=\frac{V+W}{2D},\quad W=(V^2+4Ds)^{1/2}.
\label{wdef}
\eeq
The solution obeying the boundary conditions (\ref{pbound}) reads
\beq
\w P(x,s)=\left\{\matrix{
A(r_1\,\e^{r_1x}-r_2\,\e^{r_2x})\hfill&(0<x<\ell),\hfill\cr\cr
B(\e^{r_1(x-L)}-\e^{r_2(x-L)})\hfill&(\ell<x<L).\hfill
}\right.
\label{psol}
\eeq
Finally, the constants $A$ and $B$ are determined
from the matching conditions (\ref{pmatch}):
\beq
\matrix{
A=\frac{\e^{-r_1\ell-r_2L}-\e^{-r_2\ell-r_1L}}
{D(r_2-r_1)(r_2\,\e^{-r_1L}-r_1\,\e^{-r_2L})},\hfill\cr
B=\frac{r_2\,\e^{-r_1\ell}-r_1\,\e^{-r_2\ell}}
{D(r_2-r_1)(r_2\,\e^{-r_1L}-r_1\,\e^{-r_2L})}.\hfill\cr}
\label{pcsts}
\eeq

The survival probability of the particle at time $t$,
\beq
S(t)=\int_0^L P(x,t)\,\d x,
\eeq
is nothing but the probability that the absorption time $T$ is larger than $t$:
\beq
S(t)=\int_t^\infty\rho(T)\,\d T.
\eeq
We have therefore, in Laplace space
\beq
\w S(s)=\int_0^L\w P(x,s)\,\d x,\quad
\w\rho(s)=\mean{\e^{-sT}}=1-s\w S(s).
\eeq
The solution (\ref{psol}), (\ref{pcsts}) yields after some algebra
\beq
\w S(s)=\frac{r_2(\e^{-r_1\ell}-\e^{-r_1L})-r_1(\e^{-r_2\ell}-\e^{-r_2L})}
{Dr_1r_2(r_2\,\e^{-r_1L}-r_1\,\e^{-r_2L})},
\eeq
and
\beq
\w\rho(s)
=\frac{r_2\,\e^{-r_1\ell}-r_1\,\e^{-r_2\ell}}{r_2\,\e^{-r_1L}-r_1\,\e^{-r_2L}}.
\label{rhogen}
\eeq

In the following we restrict the analysis to the limiting case
\beq
\ell\to0,
\eeq
where the particle starts from the immediate vicinity of the reflecting
endpoint.
The result (\ref{rhogen}) simplifies to
\beq
\w\rho(s)=\frac{r_2-r_1}{r_2\,\e^{-r_1L}-r_1\,\e^{-r_2L}},
\eeq
i.e., explicitly,
\beq
\w\rho(s)=\frac{2W\,\e^{VL/(2D)}}{(W+V)\,\e^{WL/(2D)}+(W-V)\,\e^{-WL/(2D)}},
\label{rhohat}
\eeq
where $W$ has been defined in (\ref{wdef}).

The moments of the absorption time $T$ can be derived by expanding
the result (\ref{rhohat}) as a power series in $s$,
as $\w\rho(s)=1-\mean{T}s+\mean{T^2}s^2/2+\cdots$
We thus obtain
\beq
\matrix{
\mean{T}=\frac{1}{V^2}(VL-D+D\e^{-VL/D}),\hfill\cr
\mean{T^2}=\frac{1}{V^4}(V^2L^2-4D^2+2D(3VL+D)\e^{-VL/D}\hfill\cr
{\hskip 18truemm}+2D^2\e^{-2VL/D}),\hfill}
\eeq
so that the reduced variance of the absorption time $T$,
\beq
K_T=\frac{\var{T}}{\mean{T}^2}=\frac{\mean{T^2}}{\mean{T}^2}-1,
\eeq
reads
\beq
K_T=D\frac{2VL-5D+4(VL+D)\e^{-VL/D}+D\e^{-2VL/D}}{(VL-D+D\e^{-VL/D})^2}.
\eeq
The above results have different kinds of behaviour in the following cases.

\noindent$\bullet$ {\it Ballistic phase} ($V>0$).
In this case, the drift brings the particle toward the absorbing endpoint.
The mean absorption time
\beq
\mean{T}\approx\frac{L}{V}-\frac{D}{V^2},
\label{btbal}
\eeq
grows linearly with $L$, according to a ballistic law with velocity $V$.
Note the negative correction due to diffusion.
Fluctuations of the absorption time around its mean are asymptotically Gaussian.
Their reduced variance,
\beq
K_T\approx\frac{2D}{VL},
\label{bkbal}
\eeq
falls off as $1/L$.

\noindent$\bullet$ {\it Diffusive point} ($V=0$).
This special case corresponds to pure diffusion.
The expression (\ref{rhohat}) simplifies to
\beq
\w\rho(s)=\frac{1}{\cosh((s/D)^{1/2}L)}.
\eeq
The mean absorption time
\beq
\mean{T}=T_0\equiv\frac{L^2}{2D}
\label{btc}
\eeq
defines the diffusive time scale $T_0$, which grows quadratically with $L$.
The Laplace transform can be inverted explicitly in this case.
The dimensionless ratio $\tau=T/T_0$ has a non-trivial distribution:
\beq
\rho(\tau)=\frac{\pi}{2}\sum_{k=0}^\infty(-1)^k(2k+1)\exp(-(2k+1)^2\pi^2\tau/8),
\eeq
with moments $\mean{\tau}=1$ (by construction),
$\mean{\tau^2}=5/3$, $\mean{\tau^3}=61/15$, and so on.
We have thus
\beq
K_T=\mean{\tau^2}-1=\frac{2}{3}.
\label{bkc}
\eeq

\noindent$\bullet$ {\it Activated phase} ($V<0$).
In this case, the drift brings the particle toward the reflecting endpoint.
It is therefore very improbable that the particle sits by chance
near the absorbing endpoint.
The absorption mechanism is therefore activated.
The mean absorption time,
\beq
\mean{T}\approx\frac{D\,\e^{\abs{V}L/D}}{V^2},
\label{bta}
\eeq
is found to grow exponentially with the length $L$.
The corresponding activation energy per unit length reads
\beq
a=\frac{\abs{V}}{D}.
\label{btadef}
\eeq
The distribution of the absorption time
can be checked to be asymptotically exponential.
In particular, the reduced variance,
\beq
K_T\approx1-\frac{2(VL+3D)}{D}\e^{-\abs{V}L/D},
\label{bk}
\eeq
converges exponentially fast to the limiting value unity,
characteristic of the exponential distribution.

\noindent$\bullet$ {\it Critical phase} ($V$ small, $L$ large).
For a large length $L$, the distribution of the absorption time
exhibits a finite-size scaling form in a narrow interval of $V$
around the diffusive point $V=0$, whose width scales as $1/L$,
where the dynamics interpolates between
the ballistic and the activated phases.
Let us introduce the dimensionless finite-size scaling variable
\beq
z=\frac{VL}{D},
\label{zeddef}
\eeq
which is twice the P\'eclet number introduced e.g. in \cite{redner}.
The moments of the absorption time scale as
\beq
\matrix{
\mean{T}=T_0\;\frac{2(z-1+\e^{-z})}{z^2},\hfill\cr
\mean{T^2}=T_0^2\;\frac{4(z^2-4)+8(3z+1)\,\e^{-z}+8\,\e^{-2z}}{z^4}.\hfill}
\label{fssf}
\eeq
The reduced variance therefore depends continuously on $z$ according to
\beq
K_T=\frac{2z-5+4(z+1)\,\e^{-z}+\e^{-2z}}{(z-1+\e^{-z})^2}.
\label{fssg}
\eeq

\section*{Appendix C. Mean hitting time for $N$ two-level systems}
\setcounter{equation}{0}
\def\theequation{C.\arabic{equation}}

In this appendix we consider an assembly of $N$ independent,
albeit not identical two-level systems,
described as spins $s_n=\pm1$ ($n=1,\dots,N$).
The spins are flipped according to independent Markov processes whose rates
\beq
w(s_n=+1\to s_n=-1)=w(s_n=-1\to s_n=+1)=\o_n
\eeq
depend on the label $n$ in an arbitrary fashion.
The stationary state of this dynamical process is an equilibrium state
where the $2^N$ configurations $\C=\{s_1,\dots,s_N\}$ are equally probable.
The above dynamics indeed obeys detailed balance with respect to
the uniform measure.

Let $\C_t$ denote the configuration of the system at time $t$,
starting from a given random initial configuration $\C_0$,
For a given stochastic history of the system,
we introduce the hitting time
\beq
T=\min\{t\st\C_t=\C_\star\},
\eeq
defined as the first time the system visits a given configuration of reference,
say $\C_\star=\{+1,\dots,+1\}$.

It will be sufficient for our purpose
to evaluate the mean hitting time $\mean{T}$,
averaged both over the random initial configuration $\C_0$,
chosen with the uniform equilibrium measure,
and over the stochastic dynamics.
Along the lines of the introduction of \cite{redner},
it is advantageous to consider simultaneously the probability
\beq
P_{\C,\C_0}(t)=\prob\{\C_t=\C\st\C_0\}
\eeq
for the system to be in configuration $\C$ at time $t$,
knowing that it was in configuration $\C_0$ at time $0$, and the probability
\beq
F_{\C,\C_0}(t)=\prob\{\C_t=\C,\;\C_t'\ne\C\;\hbox{for all}\;0<t'<t\st\C_0\}
\eeq
for the system to be in configuration $\C$ for the first time at time $t$,
again knowing that it was in configuration $\C_0$ at time $0$.
These two quantities are related by the integral equation
\beq
P_{\C,\C_0}(t)=\int_0^tF_{\C,\C_0}(t')P_{\C,\C}(t-t')\,\d t'.
\eeq
Let us average this formula with respect to
the initial configuration $\C_0$, chosen with the uniform equilibrium measure.
The left-hand side equals $1/2^N$,
for all configurations $\C$ and all times $t>0$.
The return probability $P_{\C,\C}(t-t')=R(t-t')$ is also independent of $\C$.
As a consequence, the average over $\C_0$ of $F_{\C,\C_0}(t')$
defines some average first-passage probability $F(t')$,
which does not depend on $\C$ either.
The return probability $R(t)$ and the first-passage probability $F(t)$
obey the convolution equation
\beq
\frac{1}{2^N}=\int_0^tF(t')R(t-t')\,\d t',
\eeq
i.e., in Laplace space, with the notation (\ref{lapdef}),
\beq
\frac{1}{2^N}=\w F(s)\w R(s).
\label{fr}
\eeq

In the present case of $N$ independent spins, the return probability factorises
as
\beq
R(t)=\prod_{n=1}^Nr_n(t),
\eeq
where
\beq
r_n(t)=\prob\{s_n(t)=s_n(0)\}=\half(1+\mean{s_n(t)s_n(0)}).
\eeq
The temporal correlation function of each spin variable
exhibits a pure exponential decay at equilibrium:
\beq
\mean{s_n(t)s_n(0)}=\e^{-2\o_nt},
\eeq
so that
\beq
R(t)=\frac{1}{2^N}\prod_{n=1}^N(1+\e^{-2\o_nt}).
\eeq
The return probability tends toward the limit $R(\infty)=1/2^N$, as it should.
Its Laplace transform therefore has the following behaviour as $s\to0$:
\beq
\w R(s)=\frac{1}{2^N}\left(\frac{1}{s}+C+\cdots\right),
\eeq
where the finite part $C$ is given by the convergent integral
\beq
C=\int_0^\infty\left(\prod_{n=1}^N(1+\e^{-2\o_nt})-1\right)\,\d t.
\eeq
The mean hitting time under consideration is just the mean value
of the first-passage time:
\beq
\mean{T}=\int_0^\infty t\,F(t)\,\d t=-\frac{\d\w F}{\d s}(s=0).
\eeq
Expanding (\ref{fr}) to first order in $s$, we obtain $\mean{T}=C$,
hence our final result:
\beq
\mean{T}=\int_0^\infty\left(\prod_{n=1}^N(1+\e^{-2\o_nt})-1\right)\,\d t.
\label{tint}
\eeq
Expanding the product and integrating the exponentials term by term, we obtain
\beq
\mean{T}=\half\sum_I{\frac{1}{\textstyle\sum_{n\in I}\o_n}}.
\label{tsum}
\eeq
In this expression $I$ runs over the $2^N-1$ non-empty subsets
of $\{1,\dots,N\}$.
We thus have for the first few values of $N$:
\beq
\matrix{
N=1:\quad\mean{T}=\frac{1}{2\o_1},\hfill\cr
N=2:\quad\mean{T}=\frac{1}{2}\Bigl(\frac{1}{\o_1}
+\frac{1}{\o_2}+\frac{1}{\o_1+\o_2}\Bigr),\hfill\cr
N=3:\quad\mean{T}=\frac{1}{2}
\Bigl(\frac{1}{\o_1}+\frac{1}{\o_2}+\frac{1}{\o_3}\hfill\cr
{\hskip 29truemm}+\frac{1}{\o_1+\o_2}+\frac{1}{\o_1+\o_3}\hfill\cr
{\hskip 29truemm}+\frac{1}{\o_2+\o_3}+\frac{1}{\o_1+\o_2+\o_3}\Bigr).\hfill}
\eeq

The situation of interest in the body of this paper
is where the flipping rates read $\o_n=\exdy^n$ (see (\ref{odef})),
where $\exdy$ assumes any value in the range $0<\exdy<1$.
The following special cases can be worked out more explicitly.

\noindent$\bullet$ {\it Uniform case} ($\exdy=1$).
In this case, all the flipping rates are equal ($\o_n=1$).
Equations (\ref{tint}) and (\ref{tsum}) simplify to
\beq
\mean{T}=\int_0^\infty\left((1+\e^{-2t})^N-1\right)\,\d t
=\half\sum_{m=1}^N\bin{N}{m}\frac{1}{m},
\eeq
where $m$ is nothing but the number of elements of the set $I$
entering (\ref{tsum}).
We thus obtain the estimate
\beq
\mean{T}\approx\frac{2^N}{N}.
\label{tx1}
\eeq

\noindent$\bullet$ {\it Binary case} ($\exdy=1/2$).
The problem also simplifies in this case, where $\o_n=1/2^n$.
Consider indeed (\ref{tsum}).
The denominator of the generic term can be recast as
\beq
\sum_{n\in I}\o_n=\frac{1}{2^N}\sum_{n\in I}2^{N-n},
\eeq
The sum in the right-hand side is nothing but the binary-digit expansion
of a generic integer $M=1,\dots,2^N-1$.
Therefore
\beq
\mean{T}=2^{N-1}\sum_{M=1}^{2^N-1}\frac{1}{M}.
\eeq
We thus obtain the estimate
\beq
\mean{T}\approx 2^N\,\frac{N\ln 2}{2}.
\label{txdemi}
\eeq

This particular value demarcates two different kinds of behaviour.
The change of variable $\zeta=\ln(2t)/\lambda$ in (\ref{tint}),
with the notation $\lambda=\abslnexdy$, indeed yields
\beq
\mean{T}=\frac{\lambda}{2}\int_{-\infty}^{\infty}
\left(\prod_{n=1}^N\left(1+\exp(-\e^{(\zeta-n)\lambda})\right)
-1\right)\e^{\zeta\lambda}\,\d\zeta.
\eeq
The factors in the product are approximately equal to 2
when the difference $\zeta-n$ is very negative, and to 1 when it is very
positive.
As a consequence, with exponential accuracy, we obtain
\beq
\matrix{
\mean{T}\sim\max\,(2^N,\;1/\exdy^N)\hfill\cr
{\hskip 5.3truemm}\sim\left\{\matrix{2^N\quad\hfill&(1/2<\exdy<1),\hfill\cr
1/\exdy^N\quad\hfill&(0<\exdy<1/2).\hfill}\right.
}
\label{texp}
\eeq
These estimates are exact up to $\exdy$-dependent prefactors,
as we now successively show for $1/2<\exdy<1$ and for $0<\exdy<1/2$.

\noindent$\bullet$ {\it Entropic phase} ($1/2<\exdy<1$).
In this phase, where the rates $\o_n$ exhibit a rather mild dependence on $n$,
the exponential estimate $\mean{T}\sim 2^N$ has an entropic origin.
Just as in (\ref{tentex}),
it scales as the ratio between the initial phase-space volume
(all the $2^N$ configurations) and the final one
(one single configuration of reference).

A more accurate estimate of $\mean{T}$ goes as follows.
Rewriting each factor of the product entering (\ref{tint}) as
\beq
1+\e^{-2t\exdy^n}=2\,\e^{-t\exdy^n}\cosh(t\exdy^n),
\eeq
we obtain
\beq
\mean{T}\approx 2^N A(\exdy),
\label{ta}
\eeq
with
\beq
A(\exdy)=\int_0^\infty\e^{-t\exdy/(1-\exdy)}\prod_{n=1}^\infty\cosh(t\exdy^n)
\,\d t,
\eeq
where both the product and the integral are convergent.
The series expansion of the amplitude $A(\exdy)$
as $\exdy\to1$ can be derived
by expanding the infinite product as a power series in $t$
and integrating term by term.
We thus obtain
\beq
\matrix{
A(\exdy)=(1-\exdy)+\frac{3}{2}(1-\exdy)^2\hfill\cr
{\hskip 12truemm}+\frac{5}{2}(1-\exdy)^3+\frac{19}{4}(1-\exdy)^4+\cdots}
\eeq
The amplitude $A(\exdy)$ vanishes linearly as $\exdy\to1$,
so that (\ref{tx1}) and (\ref{ta}) are compatible.
The result (\ref{txdemi}) suggests that $A(\exdy)$ diverges
linearly as $\exdy\to1/2^+$:
\beq
A(\exdy)\approx\frac{\ln 2}{4(\exdy-1/2)}.
\eeq
Finally, we have $A(\exdy)=1$ for $\exdy\approx0.6396$.

\noindent$\bullet$ {\it Slow phase} ($0<\exdy<1/2$).
In this phase, the scale of the hitting time is given by the slowest
time scale of the system: $\mean{T}\sim1/\o_N=1/\exdy^N$.

A more accurate estimate of $\mean{T}$ can be derived as follows.
Setting $n=N-j$ in (\ref{tsum}), we obtain
\beq
\mean{T}\approx\frac{B(\exdy)}{\exdy^N},
\eeq
where
\beq
B(\exdy)=\half\sum_J\frac{1}{\textstyle\sum_{j\in J}\exdy^{-j}},
\eeq
and where $J$ runs over all the non-empty subsets of the integers
$\{0,1,2,\dots\}$, i.e., explicitly
\beq
\matrix{
B(\exdy)=\half(1+2\exdy+3\exdy^2+7\exdy^3+11\exdy^4\hfill\cr
{\hskip 20truemm}+25\exdy^5+44\exdy^6+94\exdy^7+\cdots).\hfill}
\eeq
The result (\ref{txdemi}) suggests that the amplitude $B(\exdy)$
diverges linearly as $\exdy\to1/2^-$:
\beq
B(\exdy)\approx\frac{\ln 2}{4(1/2-\exdy)}.
\eeq
Finally, we have $B(\exdy)=1$ for $\exdy\approx0.2662$.

\begin{figure}[!t]
\begin{center}
\includegraphics[angle=90,height=5truecm]{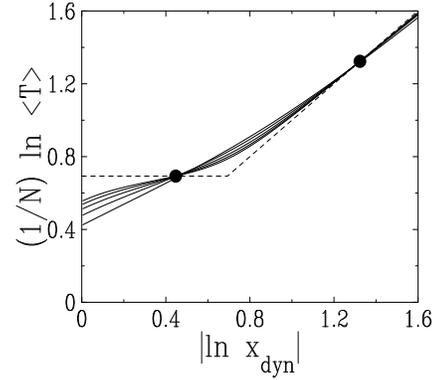}
\caption{\small
Plot of $(1/N)\ln\mean{T}$ against $\abslnexdy$,
illustrating the behaviour of $\mean{T}$ in the various phases.
Full lines: Data for $N=6$, 10, 14, 18, and 22.
Dashed straight lines: exponential estimates (\ref{texp}).
Full symbols: values of $\exdy$ where $A(\exdy)=1$ or $B(\exdy)=1$ (see text).}
\label{figar}
\end{center}
\end{figure}

Figure~\ref{figar} shows a plot of $(1/N)\ln\mean{T}$ against $\abslnexdy$,
for system sizes $N$ ranging from 6 to 22.
The data exhibit a sharper and sharper crossover
between both exponential estimates (\ref{texp}), shown as dashed lines.
The data for all the values of $N$ intersect the dashed lines
very near the theoretical values
$\exdy\approx0.2662$ and $\exdy\approx0.6396$,
where the amplitudes $B(\exdy)$ and $A(\exdy)$ are equal to unity.


\begin{thebibliography}{99}

\bibitem{0}
Anita Mehta, in {\it Granular Physics} (Cambridge University Press, Cambridge,
2007).

\bibitem{comp}
see, e.g., N. Boccara, {\it Modeling Complex Systems} (Springer, Berlin, 2004).

\bibitem{sidjamming}
E.I. Corwin, H.M. Jaeger, and S.R. Nagel, Nature {\bf 435}, 1075 (2005).

\bibitem{sid}
E.R. Nowak, J.B. Knight, E. Ben-Naim, H.M. Jaeger and S.R. Nagel, Phys. Rev.
{\bf E 57}, 1971 (1998).

\bibitem{edwards}
S.F. Edwards, in {\it Granular Matter: An Interdisciplinary Approach},
edited by A. Mehta (Springer, New York, 1994).

\bibitem{bob}
T.S. Majumdar and R.P. Behringer, Nature {\bf 435}, 1079 (2005).

\bibitem{bridges}
Anita Mehta, G.C. Barker and J.M. Luck, J. Stat. Mech. P10014 (2004).

\bibitem{I}
P.F. Stadler, A. Mehta, and J.M. Luck, Adv. Complex Systems {\bf 4}, 429
(2001).

\bibitem{II}
P.F. Stadler, J.M. Luck, and A. Mehta, Europhys. Lett. {\bf 57}, 46 (2002).

\bibitem{III}
A. Mehta and J.M. Luck, J. Phys. A {\bf 36}, L365 (2003).

\bibitem{IV}
J.M. Luck and A. Mehta, Eur. Phys. J. B {\bf 35}, 399 (2003).

\bibitem{sconf}
J. J\"ackle, Phil. Mag. B {\bf 44}, 533 (1981);
R.G. Palmer, Adv. Phys. {\bf 31}, 669 (1982).

\bibitem{miguel}
M. M\'ezard, G. Parisi, and M.A. Virasoro, {\it Spin Glass Theory and Beyond}
(World Scientific, Singapore, 1987).

\bibitem{thooft}
G. 't Hooft, in {\it Recent developments in gauge theories},
edited by G. 't Hooft et al. (Plenum, New York, 1980).

\bibitem{avalanches}
G.C. Barker and Anita Mehta, Phys. Rev. E {\bf 53}, 5704 (1996);
Phys. Rev. E {\bf 61}, 6765 (2000).

\bibitem{compcoop}
Anita Mehta, J.M. Luck, J.M. Berg, and G.C. Barker, J. Phys. Condens. Matter
{\bf 17}, S2657 (2005).

\bibitem{gl}
C. Godr\`eche and J.M. Luck, J. Phys. Condens. Matter {\bf 17}, S2573 (2005).

\bibitem{sidnature}
A.J. Liu and S.R. Nagel, Nature {\bf 396}, 21 (1998).

\bibitem{johannes}
J. Berg and Anita Mehta, Phys. Rev. E {\bf 65}, 031305 (2002).

\bibitem{br}
R.L. Brown and J.C. Richards, {\it Principles of Powder Mechanics} (Pergamon,
New York, 1966).

\bibitem{redner}
S. Redner, {\it A Guide to First-Passage Processes} (Cambridge University
Press, Cambridge, 2001).

\bibitem{feller}
W. Feller, {\it An Introduction to Probability Theory and its Applications}
(Wiley, New-York, 1966).

\end{thebibliography}
\end{document}